\documentclass[11pt,oneside,letterpaper]{article}
\usepackage{amssymb}
\usepackage{amsmath}
\usepackage[dvips]{graphicx}
\graphicspath{{./fig/}}
\usepackage{setspace}
\usepackage{amsfonts}
\usepackage{fancyhdr}
\usepackage{xcolor}
\usepackage{graphicx}
\usepackage{rotating}
\usepackage{comment}
\usepackage{color}
\usepackage{cite}
\usepackage{braket}
\usepackage{overpic}

\usepackage[export]{adjustbox}

\definecolor{darkgreen}{rgb}{0,0.5,0}
\definecolor{darkblue}{rgb}{0,0,0.6}
\definecolor{purple}{rgb}{0.4,.2,0.7}

\newcommand{\be}{\begin{equation}}
\newcommand{\ee}{\end{equation}}

\usepackage{cancel}

\usepackage[normalem]{ulem}
\usepackage[shortlabels]{enumitem}

\usepackage[colorlinks=true,citecolor=darkgreen,linkcolor=black,urlcolor=purple]{hyperref}
\usepackage{sidecap}
\usepackage{pdfsync}

\makeatletter
\newcommand*{\defeq}{\mathrel{\rlap{%
                     \raisebox{0.3ex}{$\m@th\cdot$}}%
                     \raisebox{-0.3ex}{$\m@th\cdot$}}%
                     =} 
\makeatother

\DeclareMathOperator{\Tr}{Tr}
\def\be{\begin{eqnarray}}
\def\ee{\end{eqnarray}}

\newcommand{\lan}{\langle}
\newcommand{\ran}{\rangle}
\newcommand{\tr}{\textrm{Tr}\,}

\newcommand{\ttbar}{$T\overline{T}$ }

\newcommand{\bea}{\begin{eqnarray}}
\newcommand{\eea}{\end{eqnarray}}
\def\ben{\begin{equation}}
\def\een{\end{equation}}

     \let\r=v

\def\be{\begin{equation}}
\def\ee{\end{equation}}
\def\ba{\begin{array}}
\def\ea{\end{array}}

\def\mO{{\cal O}}

\def\ba#1\ea{\begin{align}#1\end{align}}
\def\bs#1\es{\begin{split}#1\end{split}}

\allowdisplaybreaks  

\numberwithin{equation}{section}

\def \be {\begin{equation}}
\def \ee {\end{equation}}
	
\def \JM#1 {{\color{blue}  JM: #1 }}
\def \AAl#1 {{\color{red}  AA: #1 }}

\begin{document}
\onehalfspacing

\begin{center}

~
\vskip5mm

{\LARGE  {\ttbar  and the black hole interior }}

\vskip10mm

Shadi Ali Ahmad{${}^{1}$}, Ahmed Almheiri{${}^{1,2}$}, Simon Lin{${}^{2}$}
\vskip5mm

{${\ }^{1}$}{\it Center for Cosmology and Particle Physics, New York University, New York, NY 10003, USA}\\
\vskip5mm
{${\ }^{2}$}{\it  New York University Abu Dhabi, Abu Dhabi, P.O. Box 129188, United Arab Emirates }  
\vskip5mm

\href{mailto:shadiraliahmad@gmail.com}{\texttt{shadiraliahmad@gmail.com}},
\href{mailto:almheiri@nyu.edu}{\texttt{almheiri@nyu.edu}},
\href{mailto:simonlin@nyu.edu}{\texttt{simonlin@nyu.edu}}

\end{center}

\vspace{4mm}

\begin{abstract}
\noindent

There is ample evidence that the bulk dual of a \ttbar deformed holographic CFT is a gravitational system with a finite area cutoff boundary. For states dual to black holes, the finite cutoff surface cannot be moved beyond the event horizon. We overcome this by considering an extension of the \ttbar deformation with a boundary cosmological constant and a prescription for a sequence of flows that successfully pushes the cutoff boundary past the event horizon and arbitrarily close to the black hole singularity. We show how this sequence avoids the complexification of the deformed boundary energies. The approach to the singularity is reflected on the boundary by the approach of the deformed energies to an accumulation point in the limit of arbitrarily large distance in deformation space. We argue that this sequence of flows is automatically implemented by the gravitational path integral given only the values of the initial ADM charges and the area of the finite cutoff surface, suggesting a similar automatic boundary mechanism that keeps all the deformed energies real at arbitrary values of the deformation parameter. This leads to a natural definition of a deformed boundary canonical ensemble partition function that sums over the entire spectrum and remains real for any value of the deformation parameter. We find that this partition function displays Hagedorn growth at the scale set by the deformation parameter, which we associate to the region near the inner horizon in the bulk dual. 
 \end{abstract}

\pagebreak
\pagestyle{plain}

\setcounter{tocdepth}{2}
{}
\vfill
\tableofcontents

\newpage

\section{Introduction}

Barring a non-perturbative definition of probes behind the event horizon, the black hole interior will remain a mysterious place. In holography, non-perturbative effects are suppressed at the asymptotic boundary by the boundary conditions on the metric and the bulk fields, which permits a non-perturbative definition of bulk operators using the extrapolate dictionary
\begin{align}
	\lim_{r \rightarrow \infty}\phi(r,x)r^{2 \Delta} = \mO(x).
\end{align}
This level of control is not available for regions away from the asymptotic boundary. The situation is even worse for operators behind  horizons due to the inability to causally ``check'' how the operators act. To overcome these challenges, it seems necessary to shrink the gap between the boundary and the interior.

A way of moving the boundary to a finite location in the bulk was proposed by McGough, Mezei and Verlinde (MMV) \cite{McGough:2016lol} in the context of AdS$_3$/CFT$_2$. They conjectured that the family of holographic CFTs\footnote{It is important to note that the deformation can be applied to any two-dimensional QFT~\cite{Cavaglia2016}.} generated by the double trace flow \cite{Smirnov2017,Cavaglia:2016oda}
\begin{align}
    \partial_\lambda S_\lambda=  4 \int \! d^2z \,  \left( T \bar{T} - \Theta^2 \right) \label{flow}
\end{align}
corresponds to moving the holographic boundary into the bulk. This flow equation can be solved owing to  the special factorization property $\lan T \bar{T}\ran = \lan T \ran \lan\bar{T}\ran - \lan \Theta\ran^2$ proven by Zamolodchikov for any two-dimensional quantum field theory \cite{Zamolodchikov2004}. MMV motivate their holographic interpretation by matching properties of the finite cutoff bulk to those in the flowed holographic CFT. For instance, they show that the deformed energy eigenvalues of primary CFT states which take the form
\begin{align}
	E_\lambda = \frac{1}{4\lambda} \left( 1 - \sqrt{1 - 8 \lambda E_{\lambda = 0} + 64 \pi^2 J^2 \lambda^2} \right),
 \label{2ddeformedenergy}\end{align}
where $E_\lambda$ is the energy at $\lambda$ along the flow, coincides with the (renormalized) quasi-local Brown-York (BY) energy \cite{Brown:1992br} computed at the finite cutoff boundary
\begin{align}
	E = \int \sqrt{-h} \ \tilde{T}_{\mu \nu}\, u^\mu u^\nu, \quad \tilde{T}_{\mu \nu} \equiv -\frac{2}{\sqrt{-h}} {\frac{\delta S_\mathrm{bulk} }{\delta h^{\mu \nu}}}, \quad u^\mu = (\partial_t)^\mu \label{BYformula}
\end{align}
where $u^\mu$ is the unit normal to a set of bulk Cauchy surfaces  labelled by $t$ that  foliate the spacetime  and anchor on boundary Cauchy surfaces, and $h_{\mu \nu}$ is the induced metric on the finite boundary. The deformation parameter $\lambda$ is inversely related to the cutoff radius through $\lambda \sim G/  r_c^2$, where $2 \pi r_c$ is the size of the transverse space in the bulk. Other checks include the modification of thermodynamics and  superluminal propagation of boundary fluctuations (see also \cite{Marolf:2012dr}), as well as computations of correlation functions \cite{Cardy:2019qao,Li:2020pwa,He:2023kgq,Aharony:2023dod} and  entanglement entropy \cite{Donnelly:2018bef,Chen:2018eqk,Lewkowycz:2019xse} in \ttbar deformed theories.
One outstanding puzzle is the interpretation of the inevitable complexification of the deformed energies at sufficiently large $\lambda$, although see \cite{Araujo-Regado:2022gvw,Soni:2024aop} for a proposed role for describing the black hole interior.

The holographic \ttbar proposal has attracted a lot of attention over the past decade, and has since been generalized and applied to a variety of situations. It was extended to one fewer dimension in \cite{Gross:2019ach} by dimensional reduction of the $J = 0$ sector and matched to JT gravity at finite cutoff. This result was extended to define a full path integral of JT gravity at finite cutoff in \cite{Iliesiu:2020zld} (see also \cite{Bhattacharyya:2023gvg,Bhattacharyya:2025gvd,Griguolo:2021wgy}).
Higher dimensional versions and extensions were proposed in \cite{Hartman:2018tkw,Taylor:2018xcy, Morone:2024ffm, Babaei-Aghbolagh:2024hti, Brizio:2024arr,Tsolakidis:2024wut, Bhattacharyya:2023gvg, Bhattacharyya:2025gvd}. It was argued in \cite{Guica:2019nzm} that the \ttbar deformation does not alter the matter boundary conditions. This can be remedied by including a matter double trace deformation to impose  Dirichlet conditions on the finite cutoff boundary, as shown in \cite{Hartman:2018tkw, Heemskerk:2010hk, Faulkner:2010jy}. The works \cite{Caputa:2020lpa, Araujo-Regado:2022gvw} proposed a way of deforming a Euclidean CFT in order to prepare a state on a Cauchy slice of Lorentzian spacetime, and \cite{Araujo-Regado:2022gvw} conjectured an equivalence between the deformed CFT partition function and a gravitational wave function on that slice. Additional work in this direction and applying to the black hole interior includes \cite{Blacker:2024rje}. Furthermore, \cite{Gorbenko:2018oov,Lewkowycz:2019xse,Coleman:2021nor} provide a construction of a de Sitter dual from a sequence of generalizations of the \ttbar deformation.


In this paper we will analyze a flow generated by a $T \bar{T}+\Lambda_2$ deformation, and describe its role in moving the finite cutoff boundary into the black hole interior. This deformation is defined by the flow equation
\begin{align}
    \partial_\lambda S_\mathrm{QFT} = \int \! \! \sqrt{-\gamma} \left( T_{ab}T^{ab} - T^2 + \frac{1}{4\lambda^{2}}\right). 
\end{align}
Note that the sign of the $\Lambda_2$ term is opposite of that appearing in \cite{Gorbenko:2018oov,Lewkowycz:2019xse,Coleman:2021nor}. Using a sequence of deformations alternating between \ttbar and $T \bar{T}+\Lambda_2$, we will give a prescription that moves the holographic boundary throughout the BTZ geometry, shown schematically as
\begin{align}
\label{fig:flow}
\begin{matrix}
    \includegraphics[width=0.9\linewidth]{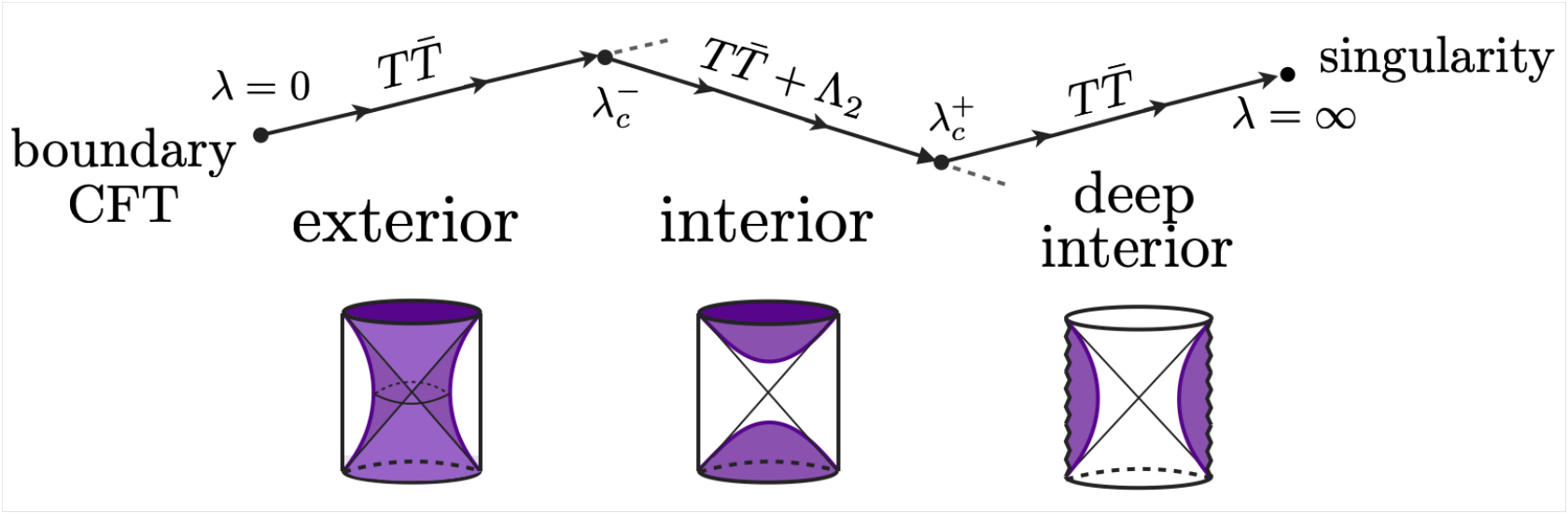} 
\end{matrix}
\end{align}
Given the boundary state in an energy and momentum eigenstate, the first step (shown on the left) is to perform the regular \ttbar deformation which pushes the cutoff boundary into the exterior of the bulk away from the asympotitc boundary. At some critical value of deformation parameter $\lambda_c^-$ we will encounter the outer horizon. The second step (shown in the middle) we turn on a boundary cosmological constant term $\Lambda_2$ in the deformation. The addition of the $\Lambda_2$ term avoids the issue of complexification of the energy and puts us in the region between outer horizon and inner horizon in the bulk. The final step (shown in the right) takes place after the deformation hits the inner horizon, as signaled by another critical value $\lambda_c^+$. To proceed we must turn off the boundary $\Lambda_2$ term. The flow can now be safely resumed and we reach the bulk region between the singularity and the inner horizon, which we term the ``deep interior''. For black holes with zero angular momentum, the region corresponding to the deep interior is absent and $\lambda_c^+=\infty$  corresponds to the conical singularity .

The paper is organized as follows: In {\bf{section \ref{sec:review}}}, we review how to define and solve \ttbar flowed theories in two ways: one using the original operator method of \cite{Smirnov2017,Cavaglia:2016oda } and the other using the generating functional method of \cite{Guica:2019nzm}. In {\bf{section \ref{sec:ttbarCC}}}, we define our generalization of the \ttbar flow that includes a cosmological constant, and proceed to solve the theory using the same methods as in the previous section and explain how to construct the desired sequence of flows. This sequence will avoid the complexification of the energies and allow us to continue increasing the deformation parameter indefinitely. We then turn to the bulk in {\bf{section \ref{sec:bulkflow}}} to motivate the \ttbar deformation and its generalization along the lines of \cite{Hartman:2018tkw, Taylor:2018xcy}. In {\bf{section \ref{sec:GPI}}}, we propose a gravitational path integral that implements the boundary \ttbar prescription and give the boundary conditions dual to fixing the initial energy, angular momentum, and the deformation parameter $\lambda$. We show how these parameters control the location of the finite cutoff surface and how they can place the cutoff surface behind the horizon of a black hole. Surprisingly, we find that the gravity path integral  avoids the complexification of the deformed energies by automatically implementing the needed sequence of flows. This motivates a similar automatic boundary mechanism, and  supports a definition of a deformed canonical partition function summing over all real deformed energies.

\section{Review: \ttbar flow}\label{sec:review}

We begin by reviewing the \ttbar proposal emphasizing essential elements for what follows. We will start with the operator approach of Smirnov and Zamolodchikov \cite{Smirnov2017} followed by the generating functional approach of Guica and Monten \cite{Guica:2019nzm}. See  \cite{He:2025ppz} for a recent review.

\subsection{Operator method}
Consider a Lorentzian CFT on the cylinder with radius $L$. The \ttbar deformation defines a one paramter family of QFTs related by the flow equation
\begin{align}
   Z_\lambda = \int \! \!  d\psi \, e^{- i S_\lambda} \rightarrow \partial_\lambda S_\lambda&=   \int \! \! dt d\theta \left( T_{ab}T^{ab} - T^2 \right), \label{originalttbar} \\
    &= 2\int \! \! dt d\theta \left( T_{tt}T_{\theta \theta} - T_{t \theta}^2 \right)
\end{align}
Incremental flow along this space of theories can be understood as adding an infinitesimal perturbation to the Hamiltonian at each step. The Feynman-Hellman theorem of time independent perturbation theory states that the first order  correction to the energy\footnote{Our convention for the relation of the partition function to the path integral is such that $Z = \tr e^{- i H (t+ i \epsilon) }$, where $\epsilon$ is a regulator.} in any eigenstate $| E, J \ran$ must satisfy
\begin{align}
	\partial_\lambda E &=\lan E,J| \partial_\lambda H | E,J \ran\\
	&= 2 \int d\theta \, \lan E,J|\left( T_{tt}T_{\theta \theta} - T_{t \theta}^2 \right) | E, J \ran \\
	&= 2 L \left( \lan T_{t t} \ran \lan T_{\theta \theta} \ran  - \lan T_{t \theta} \ran^2\right)
\end{align}
where the expectation values in the last line are in the state $| E,J\ran$.  Assuming  this state is primary, we  used the translation invariance  to implement the factorization property of the \ttbar operator and to also evaluate the spatial integral. The resulting differential equation can be recast in terms of the energy and momentum eigenvalues through 
\begin{align}
	\lan T_{t t} \ran = \frac{E}{L}, \quad \lan T_{\theta \theta} \ran = -\frac{\partial E}{\partial L}, \quad \lan T_{t \theta} \ran = -\frac{2\pi J}{L^2} ,
\end{align}
to give the first order differential equation
\begin{align}
	{1 \over 2}\partial_\lambda E = - E {\partial E \over \partial L} - {4 \pi^2 J^2 \over L^3}. \label{energyflowttbar}
\end{align}
Since the spatial direction is compact, the angular momentum is quantized and cannot flow. The trick to solving this equation is to note that the energy can be written as $E(\lambda, L) \equiv L^{-1} \times f(\lambda/L^2)$ because $L$ and $\lambda \sim L^2$ are the only two scales in the problem. This allows us to solve the above equation to get the famous result
\begin{align}
	E_\lambda = {L \over 4 \lambda} \left( 1 - \sqrt{1 - 8 \lambda E_0/L + 64 \pi^2 J^2 \lambda^2/L^4} \right). \label{defE}
\end{align}
where $E_0$ is the undeformed value of the energy at $\lambda = 0$, sometimes referred to as the initial or ``seed'' energy.

For applications of \ttbar to holography, the deformation parameter is assumed to be positive $\lambda >0$. Therefore, for any given initial $E_0$ and $J$, the deformed energy will complexify for a range of the deformation parameter bounded by the two roots of the argument of the square root
\begin{align}
    \lambda_c^\pm = {2 E_0 \pm \sqrt{4 E_0^2 - (2 \pi J)^2 }\over 2 (2 \pi J)^2}. \label{lambdac}
\end{align}
We will propose a way of evolving beyond $\lambda_c^-$ with a modified deformation which we will argue corresponds to evolving into the black hole interior; more on that in section \ref{sec:GPI}.

\begin{figure}[ht]
    \centering
    \includegraphics[width=0.49\linewidth]{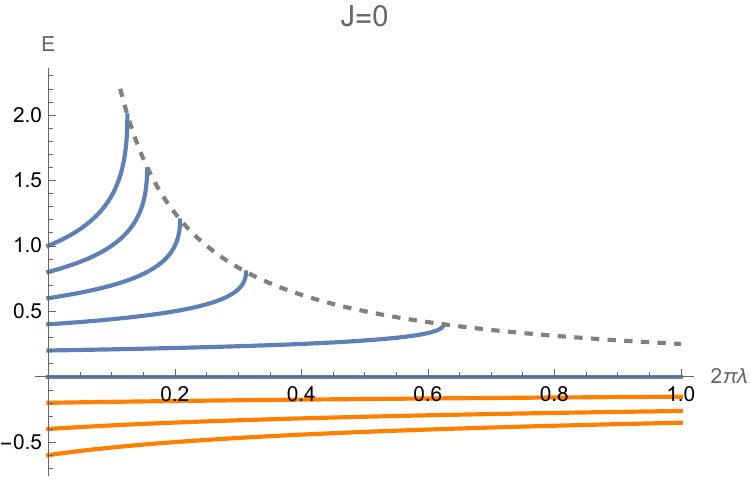}
    \includegraphics[width=0.49\linewidth]{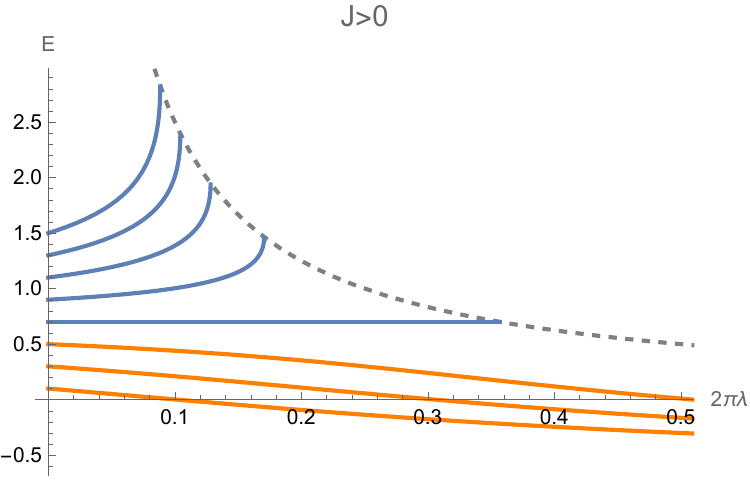}
    \caption{The flowed energy levels \eqref{defE} for various different seed energies $E_0$. We set $L=2\pi$ here. The left figure has $J=0$ and the right figure has $J=0.7$.  The blue lines are states with $E_0\ge J$, and the orange lines feature $E_0<J$. For the $E_0>J$ solutions, the energy complexifies at $\lambda_c^-$, which is indicated by the dashed curve. The $E_0<J$ solutions do not complexify.}
    \label{fig:energy_level_ext}
\end{figure}

\subsection{Generating functional method}\label{sec:review_gen_method}

Next, we review the generating functional method introduced in \cite{Guica:2019nzm}. In addition to the flow of the stress energy tensor, this method captures how the background metric of the theory is modified along the flow. Properties of the deformed metric will be important for the remainder of the paper.

There exists a large body of work on deforming CFTs by double-trace deformations and their holographic duals; a selection thereof is \cite{Aharony:2001pa, Berkooz:2002ug, Papadimitriou:2007sj, Heemskerk:2010hk, Faulkner:2010jy}. The upshot is that, in a theory with large $N$ factorization, the one-point function of an operator $\mO$ in a theory deformed by $\lambda \mO^2$ by an infinitesimal amount $\lambda$ is related to the one-point function in the undeformed theory by
\begin{align}
	\lan \mO \ran_\lambda^{J_\lambda} = \lan \mO \ran_0^{J_\lambda + \lambda \lan \mO \ran_0^{J_\lambda}}, \label{doubletrace}
\end{align}
where subscripts $\lambda$ and $0$ refer to the deformed and undeformed theories respectively. In holography, the two sides of this equation can be understood as follows. The right hand side shows that the asymptotic boundary conditions have changed from Dirichlet to a mixture of Dirichlet and Neumann. The left hand side says that this is equivalent to imposing just Dirichlet boundary conditions on a surface infinitesimally shifted from the asymptotic boundary. 


This equation relates the value of the source on the $\lambda$ surface to the required source at infinity, namely $J_\lambda = J_0 - \lambda \lan \mO \ran_0^{J_\lambda}$.  This implies that the generating functionals of the two theories must be related through
\begin{align}
	-W_\lambda[J_\lambda] = -W_0[J_\lambda + \lambda \lan \mO \ran_0^{J_\lambda}] + {\lambda \over 2} \left( \lan \mO \ran_0^{J_\lambda} \right)^2. \label{generalgenerating}
\end{align}
As discussed in \cite{Guica:2019nzm}, this relation between generating functionals follows from defining the deformed action as $S_\lambda = S_0 + {\lambda \over 2}\mO^2$, notably with an opposite sign compared to the deformation of the generating functional.\footnote{A quick way to see this is to use the on-shell expression $W[J] = S + J \mO$ on both sides of \eqref{generalgenerating}.}

The above discussion also applies to the \ttbar deformation, except it not necessary to assume large $N$ approximation if we are in a primary state where the \ttbar operator already factorizes. Therefore, the stress tensor generating functional satisfies the differential equation
\begin{gather}
    \partial_\lambda W_\lambda[\gamma^{ab}] = - \int \! \sqrt{-\gamma} \, \left( {{T}}^{a b}  {{T}}_{ab} - ({{T}}^a_a)^2 \right).\label{wflowlor}
\end{gather}
 Note that all  operators in this expression should be thought of as expectation values. Solving this equation will produce flows for the stress energy tensor and the source (here being the metric). As discussed above, the metric flow will determine how the metric in the deformed theory  is related to the one in the undeformed theory.

 This equation is solved by applying an infinitesimal variation to both sides. Using $W \equiv i \ln Z$ and
 \begin{align}
    \delta (i \ln Z) ={1 \over 2} \int \! \sqrt{-\gamma} \, T_{ab} \gamma^{ab},
\end{align} 
which is our convention for a Lorentzian QFT,\footnote{Our convention  is set by 
\begin{align}
    E = i \partial_t \ln e^{-i E t} =  i {2 \over \sqrt{-h}}  {\delta \over \delta h^{tt}} \ln Z = T_{tt}\implies T_{ab} = { {2 \over \sqrt{-h}}  {\delta \over \delta h^{ab}} (i\ln Z)}
\end{align}} 
we find the flow equation 
\begin{align}
    \partial_\lambda \left( \sqrt{-\gamma} \delta \gamma^{ab}{{T}}_{ab} \right) = -2\,  \delta \! \left( \sqrt{-\gamma} \left( {{T}}^{a b} {{T}}_{ab} - ({{T}}^a_a)^2 \right) \right). \label{variedflow}
\end{align}
This breaks up into the set of coupled flow equations for the metric and the stress tensor
\begin{align}
    \partial_\lambda \gamma_{ab} = 4 \left( {{T}}_{ab} - \gamma_{ab} {{T}} \right) \equiv 4 \hat{{{T}}}_{a b}, \quad \partial_\lambda \hat{{{T}}}_{ab} = 2 \hat{{{T}}}_{ac}\hat{{{T}}}_{b}^{\ c}, \quad \partial_\lambda (\hat{{{T}}}_{ac}\hat{{{T}}}_{b}^{\ c}) = 0. \label{guicaflow}
\end{align}
The details of this derivation can be found in \cite{Guica:2019nzm}. The general solution  is
\begin{align}
\label{eq:metricflow}
    \gamma_{ab} &= \gamma^{[0]}_{ab} + 4 \lambda \hat{{{T}}}^{[0]}_{a b} + 4 \lambda^2 \hat{{{T}}}^{[0]}_{a c}\hat{{{T}}}^{[0]}_{b d}\gamma^{[0]cd}, \\
    \hat{T}_{ab}&=  \hat{T}^{[0]}_{a b} + 2 \lambda \hat{T}^{[0]}_{a c}\hat{T}^{[0]}_{b d}\gamma^{[0]cd},
\end{align}
where $\gamma^{[0]}_{ab}$ and $T^{[0]}_{a b}$ are the initial values of the metric and stress tensor at $\lambda =0$. We are interested in applying \ttbar to a QFT defined on flat space, and hence we take  $\gamma^{[0]}_{ab}$ to describe a two dimensional flat Lorentzian metric in some gauge. In fact, we can use the gauge freedom in the tensor flow equations \eqref{guicaflow} to pick the initial metric to be the Minkowski metric $\gamma^{[0]}_{ab} = \eta_{ab}$. As shown in \cite{Guica:2019nzm}, \ttbar flow does not modify the curvature of the background geometry but it  implements a  $\lambda$-dependent coordinate transformation on the initial metric. In our case, we can write the flowed metric as a diffeomorphism  from the Minkowski metric
\begin{align}
    \gamma_{ab} &= \eta_{ab} + 4 \lambda \hat{{{T}}}^{[0]}_{a b} + 4 \lambda^2 \hat{{{T}}}^{[0]}_{a c}\hat{{{T}}}^{[0]}_{b d}\eta^{cd} \\
    &= \left( \delta_{a}^c + 2 \lambda \hat{{{T}}}^{[0]}_{a e}\eta^{e c}  \right)\eta_{c d}\left( \delta_{b}^d + 2 \lambda \eta^{d f} \hat{{{T}}}^{[0]}_{ f b}  \right) ,\label{flowedflat}
\end{align}
where the $\lambda$-dependent diffeomorphism satisfies
\begin{align}
    {dy_\lambda^c \over d x^a} = \delta_{a}^c + 2 \lambda \hat{{{T}}}^{[0]}_{a e}\eta^{c e}  . \label{lambdacoord}
\end{align}

The metric and stress tensor solutions depend on the stress tensor components $ \hat{{{T}}}^{[0]}_{a b}$ which set the initial condition for the stress tensor flow. The flowed stress tensor can be written as the initial stress tensor up to a coordinate transformation implemented on one of its indices as
\begin{align}
    \hat{T}_{ab} =  \hat{T}_{ac}^{[0]} \left(\delta_{b}^c + 2\lambda \hat{{{T}}}^{[0]}_{b e}\eta^{c e}  \right). \label{Tcoord}
\end{align}
This result is very different from the deformed energy in the operator method section above. This is because these stress tensor components are in a gauge set by the flowed metric. To recover the result of the previous section,   we need  to find the stress tensor in the appropriate Minkowski frame. While \cite{Guica:2019nzm} demonstrates through a bulk argument, we will present a purely field theory argument for the same outcome. We do this by requiring the radius of the cylinder to be fixed along the flow. Before deriving this, we first describe all the coordinate systems appearing in the problem and their corresponding stress tensors.
\begin{enumerate}
    \item Let $x^a$ be the coordinate system describing the solution \eqref{flowedflat}. The components $T_{ab}$ are those of the deformed stress tensor on the metric $\gamma_{ab}dx^a dx^b$, while $T^{[0]}_{ab}$ are the undeformed stress tensor  components using the same coordinates but in the metric $\eta_{ab}dx^a dx^b$.
    \item Let $y^a_\lambda$ be the coordinates where the metric $\gamma_{ab}$ has Minkowski form, namely which satisfies $\gamma_{ab}dx^a dx^b = \eta_{cd}dy_\lambda^c dy_\lambda^d$. The coordinate transformation between them satisfies equation \eqref{lambdacoord}. Note that there is an overall  $SO(1,1)$ gauge freedom that we will fix later. Let ${\cal{T}}_{cd}$ be the stress tensor components in the $y^c_\lambda$ coordinates. It satisfies  
    \begin{align}
    \label{T_sandwich}
        T_{ab} = \left( \delta_{a}^c + 2 \lambda \hat{{{T}}}^{[0]}_{a e}\eta^{c e}\right) {\cal{T}}_{cd} \left( \delta_{b}^d + 2 \lambda \hat{{{T}}}^{[0]}_{b f}\eta^{d f} \right)
    \end{align}
    which by \eqref{Tcoord} implies
    \begin{align}
       \hat{T}_{ad}^{[0]} \left(\delta_{b}^d + 2 \lambda \hat{{{T}}}^{[0]}_{b e}\eta^{d e}  \right) = \left( \delta_{a}^c + 2 \lambda \hat{{{T}}}^{[0]}_{a e}\eta^{c e}\right) \hat{{\cal{T}}}_{cd} \left( \delta_{b}^d + 2 \lambda \hat{{{T}}}^{[0]}_{b f}\eta^{d f} \right) \label{T0Ttilde}
    \end{align}
    \item Let $y^a_0$ be the coordinate system where the $\lambda = 0$ metric has Minkowski form and where the size of the periodic direction is equal to that in the $y^c_\lambda$ coordinate. Its relation to the $\lambda = 0$ metric is
    \begin{align}
        \eta_{ab} dx^a dx^b = \Omega^2 \eta_{cd} dy^c_0 dy^d_0. \label{undeformedmetrics}
    \end{align}
    where the (constant) conformal factor is needed to rescale the periodic direction in the $x^a$ coordinates. Let ${\cal T}^0_{cd}$ be the stress tensor on the metric $\eta_{cd} dy^c_0 dy^d_0$.\footnote{We can ignore the conformal factor $\Omega^2$ because the stress tensor is invariant under constant conformal transformations.}  It satisfies
    \begin{align}
        {\cal T}^0_{cd} = {d x^a \over dy_0^c}{d x^b \over dy_0^d} T^{[0]}_{ab}. \label{undeformedTs}
    \end{align}
Note that there is another $SO(1,1)$ gauge freedom in picking the $y^a_0$ coordinates. This  we can fix by specifying the components of the stress tensor $ {\cal T}^0_{cd}$.
\end{enumerate}

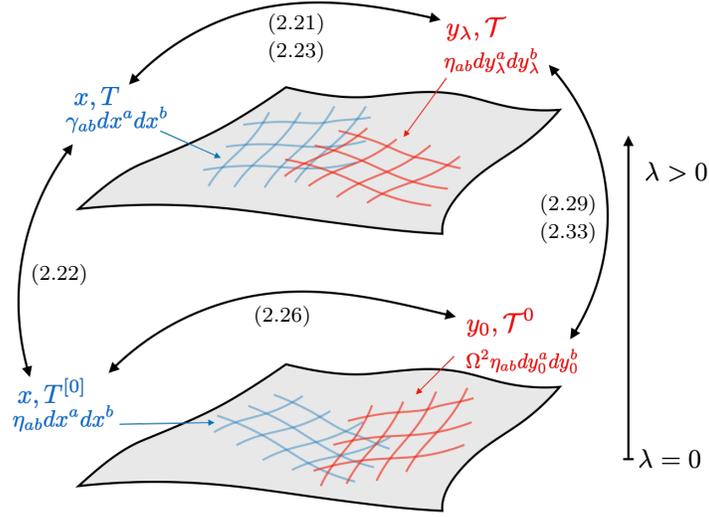
\begin{figure}[t]
    \centering
    \begin{overpic}[width=0.6\linewidth]{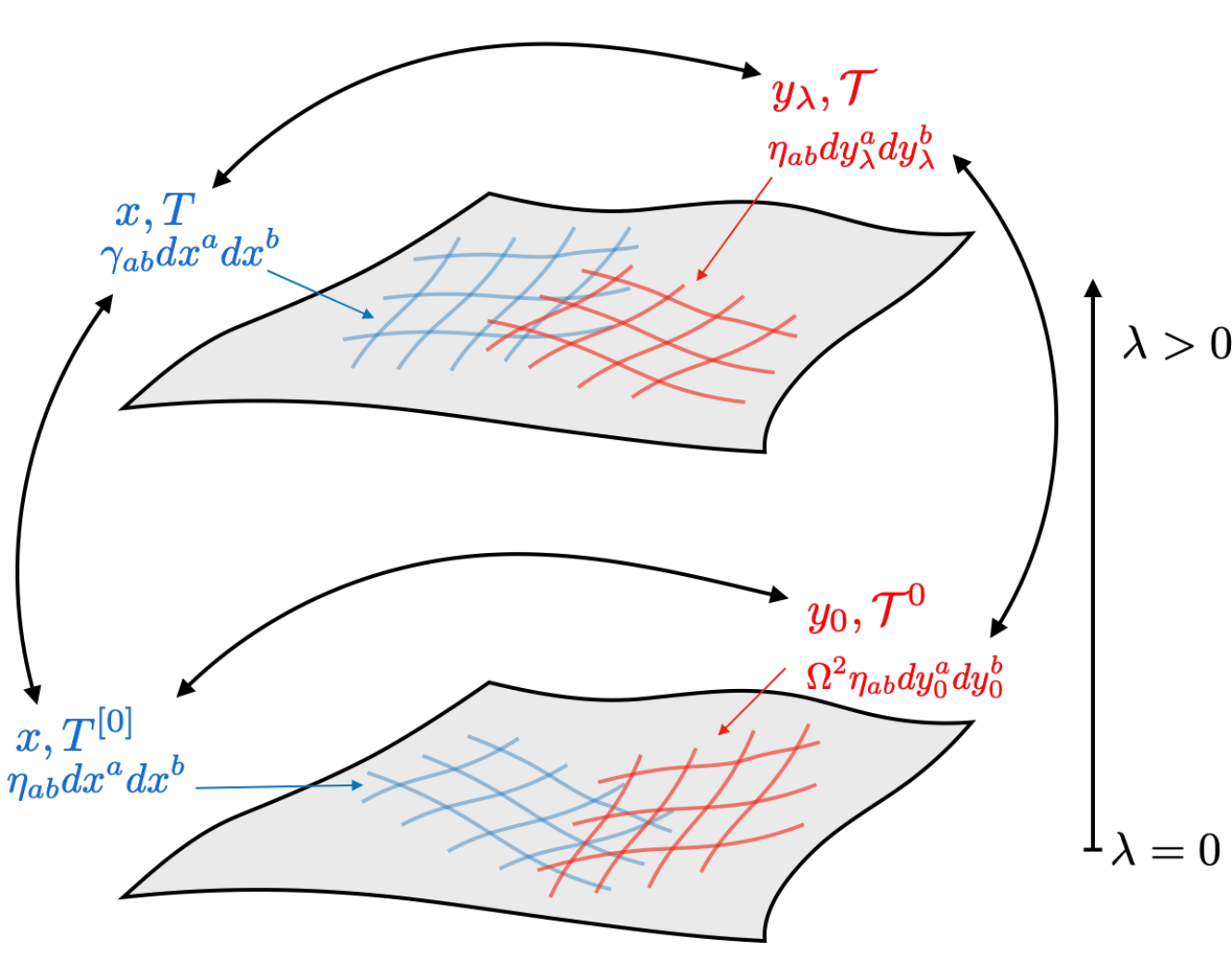}
    \put(3,35){\scriptsize \eqref{Tcoord}}
    \put(35,29){\scriptsize \eqref{undeformedTs}}
    \put(37,71){\scriptsize \eqref{lambdacoord}}
    \put(37,67){\scriptsize \eqref{T_sandwich}}
    \put(76,45){\scriptsize \eqref{dyldy0}}
    \put(76,41){\scriptsize \eqref{T0_sandwich}}
    \end{overpic}
    \caption{Illustration of the relations between the metric and  stress tensor in the coordinate systems appearing in this subsection.}
    \label{fig:map}
\end{figure}


The upshot is that the undeformed and deformed stress tensors of the operator method section are  ${\cal T}^0_{ab}$ and  ${\cal T}_{ab}$ respectively, and we need to find the transformation that relates the two. Using equations \eqref{undeformedTs} and \eqref{T0Ttilde} we have
\begin{align}
     {\cal T}^0_{cd} &= {d x^a \over dy_0^c} \left( \delta_{a}^f + 2 \lambda \hat{{{T}}}^{[0]}_{a e}\eta^{f e}\right) \hat{{\cal{T}}}_{fb}{d x^b \over dy_0^d}. \label{ToT}
\end{align}
Our strategy will to describe next how to find all the factors on the right hand side purely in terms of ${\cal T}_{cd}$, and then invert the equation to obtain ${\cal T}_{cd}$ in terms of ${\cal T}_{cd}^0$.

The first step is to relate all the coordinates above.  The first $SO(1,1)$ gauge redundancy  we encountered is removed by requiring  the length in the periodic direction to be the same in the $y_\lambda^a$ and $y_0^a$ coordinates. This translates into the requirement that paths at constant time in both coordinates satisfy
    \begin{align}
        dy^\tau_\lambda = dy^\tau_0 = 0, \quad dy^\theta_\lambda = dy^\theta_0.
    \end{align}
This fixes the relation between the infinitesimals to be
    \begin{align}
        dy^c_\lambda = \left( \delta^{c\theta}\delta_{a \theta}+ A \delta^{c \theta}\delta_{a \tau } + B \delta^{c \tau}\delta_{a \tau}   \right) dy^a_0 \rightarrow dY_\lambda = \Lambda \cdot dY_0, \label{dyldy0}
    \end{align}
expressed in matrix form on the right with $\Lambda^c_a = \delta^{c\theta}\delta_{a \theta}+ A \delta^{c \theta}\delta_{a \tau } + B \delta^{c \tau}\delta_{a \tau} $ for some $A,B$ we need to find. By using \eqref{lambdacoord} and \eqref{T0Ttilde} we find
\begin{align}
    dY_\lambda = \left( I + 2 \lambda  \eta^{-1} \hat{{{T}}}^{[0]} \right) \cdot dX = \left( I - 2 \lambda \eta^{-1} \hat{{\cal{T}}}  \right)^{-1} \cdot dX, \label{dyldx}
\end{align}
which allows us to  write the constraint \eqref{undeformedmetrics} as
\begin{align}
   \Lambda^T \cdot \left( I - 2 \lambda \eta^{-1}\hat{{\cal{T}}} \right)^T \cdot \eta \cdot \left( I - 2 \lambda \eta^{-1}\hat{{\cal{T}}} \right)\cdot \Lambda  = \Omega^2 \eta.
\end{align}
This means that $\left( I - 2 \lambda \eta^{-1}\hat{{\cal{T}}} \right)\cdot \Lambda $ must be an element of $O(1,1)$ with determinant $\Omega^2$. This completely fixes $\Lambda$ in terms of $\hat{{\cal{T}}}$ and $\lambda$. Finally, combining \eqref{dyldy0}, \eqref{dyldx}, and \eqref{ToT} we get
\begin{align}
    {\cal T}^0 &= \Lambda^T \left(1-2 \lambda \eta^{-1} \hat{{\cal{T}}} \right)^T T^{[0]} \left( 1 -2 \lambda \eta^{-1} \hat{{\cal{T}}} \right)\Lambda \\
    &= \Lambda^T  \hat{{\cal{T}}} \left( 1 -2 \lambda \eta^{-1} \hat{{\cal{T}}}  \right)\Lambda.
    \label{T0_sandwich}
\end{align}
We can now solve for ${\cal T}$ by inverting this relation and picking the solution that satisfies the boundary condition ${\cal T} \rightarrow {\cal T}^0$ as $\lambda \rightarrow 0$. The result is
\begin{align}
    {\cal T}_{tt} &= {1 \over 4 \lambda} \left( 1 - \sqrt{1 - 8 \lambda {\cal T}^0_{tt} + 16 \lambda^2 ({\cal T}^0_{tx})^2}\right), \\
    {\cal T}_{tx} &= {\cal T}_{tx}^0, \\
    {\cal T}_{xx} &={1 \over 4 \lambda} { 1 - 16 \lambda^2 ({\cal T}^0_{tx})^2-\sqrt{1 - 8 \lambda {\cal T}^0_{tt} + 16 \lambda^2 ({\cal T}^0_{tx})^2}  \over \sqrt{1 - 8 \lambda {\cal T}^0_{tt} + 16 \lambda^2 ({\cal T}^0_{tx})^2}},
\end{align}
 consistent with the results of the previous section.

Finally, we note that the metric has some interesting behavior in the limit as $\lambda \rightarrow \lambda_c^-$. Using the results above, the flowed metric determinant satisfies
\begin{align}
    \det[ \gamma] &= - \left( \det\left[ I + 2 \lambda \eta^{-1}\hat{{{T}}}^{[0]} \right] \right)^{2} ,\\
    &= - \left( \det\left[ I - 2 \lambda \eta^{-1}\hat{{\cal{T}}} \right] \right)^{-2} ,\\
    &= -\left( {2 \sqrt{1 - 8 \lambda {\cal T}^0_{tt} + 16 \lambda^2 ({\cal T}^0_{tx})^2} \over 1- 4{\cal T}^0_{tx} + \sqrt{1 - 8 \lambda {\cal T}^0_{tt} + 16 \lambda^2 ({\cal T}^0_{tx})^2}} \right)^2, \label{detgammattbar}
\end{align}
where the minus sign comes from the determinant of the Minkowski metric. The determinant of the metric is negative and ultimately degenerates as $\lambda \rightarrow \lambda_c^-$ when the numerator vanishes. The new flow introduced in the next section will allow us to evolve past $\lambda_c^-$ and change the signature of the metric.
\section{\ttbar + $\Lambda_2$}\label{sec:ttbarCC}

In this section we will solve a generalization of the \ttbar deformation that includes a  $\lambda$ dependent cosmological constant $\Lambda_2$. The modified flow equation is given by
\begin{align}
    \partial_\lambda S_\mathrm{QFT} = \int \! \! \sqrt{\gamma} \left( T_{ab}T^{ab} - T^2 + {1 \over 4 \lambda^2}\right). \label{ttbarcc}
\end{align}
This flow will be implemented after a finite amount of the standard \ttbar flow,  and so one need not worry worry about a potential divergence of the $\Lambda_2$ term. In particular, the initial conditions for this flow will be set by the final outcome of the original \ttbar flow at $\lambda_c^-$. We will provide a prescription that allows this modified flow to change the signature of the boundary metric and avoid the complexification of the deformed energies at $\lambda > \lambda_c^-$. This behavior is reminiscent of the exchange of causal character of the time and radial coordinates upon crossing a Killing horizon.

Mirroring the presentation of the previous section, we will solve this flow using the operator and generating functional methods. We begin by assuming the signature change from the previous flow which we will argue for when analyzing the metric flow using the generating functional method.
\subsection{Operator Method}\label{sec:ttbarCC_Op}
Consider the flow equation for a theory with Euclidean metric $dt^2 + d\theta^2$,
\begin{align}
    \partial_\lambda S_\lambda  = - 2 \int dt d\theta \left(   T_{tt}T_{\theta\theta} - T^2_{t\theta} - { 1 \over 8 \lambda^2}   \right). 
\end{align}
The stress tensor is defined in the usual way for a Euclidean theory\footnote{The convention here is set by 
\begin{align}
    E = - \partial_\tau \ln e^{- E \tau} =   {2 \over \sqrt{h}}  {\delta \over \delta h^{\tau \tau}} \ln Z = - T_{\tau \tau} \implies T_{ab} = {2 \over \sqrt{h}} {\delta \over \delta h^{ab}} (- \ln Z).
\end{align}}  using 
\begin{align}
     \lan T_{ab} \ran &\equiv {2 \over \sqrt{\gamma}} {\delta \over \delta \gamma^{ab}}(-\ln Z) \\
     &= {2 \over \sqrt{\gamma}} {\delta \over \delta \gamma^{ab}}(i S)
\end{align}
where the $i$ comes from the definition partition function in the initial flow defined as $Z = e^{- i S}$ (see \eqref{originalttbar}) and the $i$ is retained along the flow. In the previous section we assumed that the partition function is proportional to the time evolution operator given by $e^{-i(H - \Omega J)t}$. For reasons that will be clear once we discuss the dual description in sections \ref{sec:bulkflow} and \ref{sec:GPI}, we must take the time evolution operator in this case to be $e^{i(H - \Omega J)t}$.\footnote{A potential boundary argument we could not make precise: we start by taking $t \rightarrow - i t$ since the signature of the time coordinate flips, and then evolve using $ - i H$ since $H$ generates Lorentzian time evolution while $-iH$ generates Euclidean evolution.} This implies that the sign of the Hamiltonian is opposite that of the action. 

Since the functional derivative $ {2 \over \sqrt{\gamma}} {\delta \over \delta \gamma^{tt}} \sim - \partial_t$, its action on $\ln Z$ should give $-i E$ and so the partition function behaves like a thermal ensemble with imaginary energy.

The Feynman-Hellmann theorem once again implies 
\begin{align}
    \partial_\lambda E  &= 2 \int d\theta \, \lan E,J|\left( T_{tt}T_{\theta \theta} - T_{t \theta}^2 - { 1 \over 8 \lambda^2}\right) | E, J \ran \\
	&= 2 L \left( \lan T_{t t} \ran \lan T_{\theta \theta} \ran  - \lan T_{t \theta} \ran^2 - { 1 \over 8 \lambda^2}\right).
\end{align}
The components of the stress tensor via our convention\footnote{One check of the relative sign between $T_{tt}$ and $T_{\theta \theta}$ is the requirement that the trace of the stress tensor vanishes if we have conformal symmetry.} are given by
\begin{align}
	\lan T_{t t} \ran = {iE \over L}, \quad \lan T_{\theta \theta} \ran = i{\partial E \over \partial L}, \quad \lan T_{t \theta} \ran = {2 \pi i J \over L^2},
\end{align}
resulting in the differential equation
\begin{align}
	{1 \over 2}\partial_\lambda E = - E {\partial E \over \partial L} + {4 \pi^2 J^2 \over L^3}- {1 \over 8 \lambda^2}. \label{energyflowttbarCC}
\end{align}
As in the previous section, this is a first order differential whose solution is fixed with one boundary condition. We match onto the original \ttbar flow at $\lambda_c^-$, namely
\begin{align}
    E_{\lambda_c^-}^{T \bar{T} + 1/4\lambda^2} = E_{\lambda_c^-}^{T \bar{T}}.
\end{align}
The angular momentum remains unchanged along the flow. The solution we obtain is
\begin{align}
\label{defEint}
    E_\lambda = {L \over 4 \lambda} \left( 1 - \sqrt{-1 + 8 \lambda E_0/L - 64  \pi^2 J^2 \lambda^2 /L^4} \right).
\end{align}
This result coincides with \eqref{defE} from the previous section up to a sign flip of the argument of the square root, and hence the energy does not immediately complexify once $\lambda > \lambda_c^-$. However, it would complexify once $\lambda>\lambda_c^+$. We will revisit this later in this section and eventually argue that it corresponds to evolving past the inner horizon of a rotating BTZ black hole.

\begin{figure}[ht]
    \centering
    \includegraphics[width=0.49\linewidth]{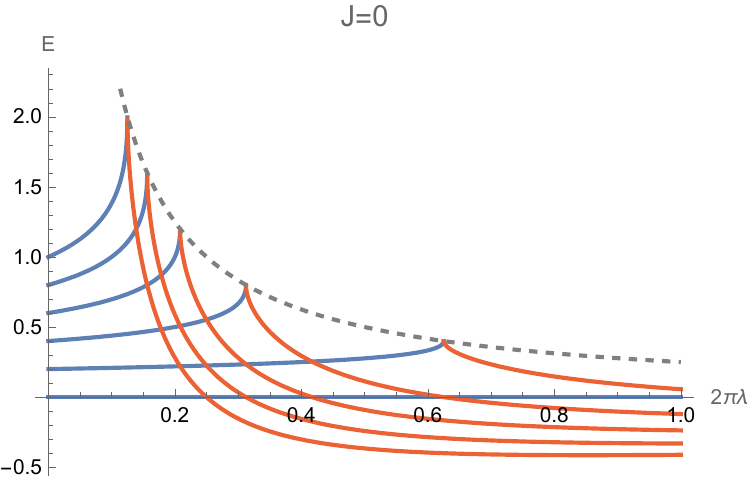}
    \includegraphics[width=0.49\linewidth]{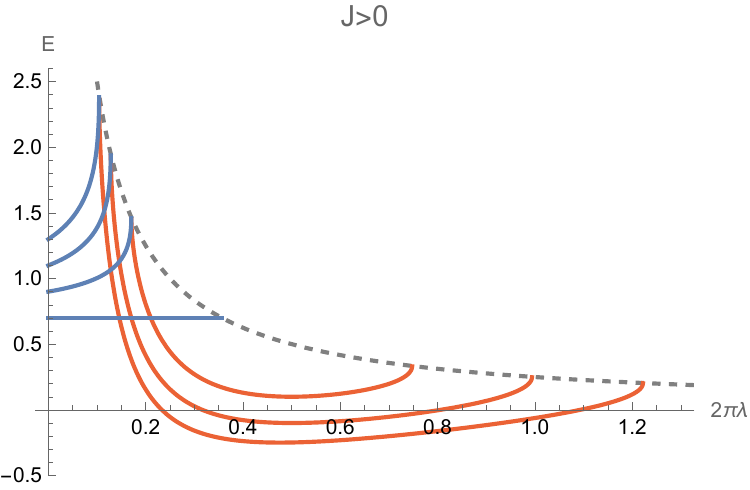}
    \caption{The extended flowed energy levels for various different seed energies $E_0$. The left figure has $J=0$ the right figure has $J=0.7$. The blue curves are the flowed energy levels \eqref{defE} in the first stage of the flow, and the red curves are the continuation of the flow given by \eqref{defEint}.  We omit the $E_0<J$ states here since they never complexify along the first flow. For the $J>0$ case, the interior energy complexifies again after reaching the second critical value $\lambda^+_c$.}
    \label{fig:energy_level_int}
\end{figure}

\subsection{Generating functional method}\label{sec:ttbar+cc:generatingfunctionalmethod}

Here we apply the generating functional method to solve the modified flow \eqref{ttbarcc}. The aim is to reproduce the flowed energy obtained from the operator method and to demonstrate the metric signature change.

The deformation to the generating functional is
\begin{align}
    \partial_\lambda W [\gamma^{ab}] =  \int \! \! \sqrt{\gamma} \left( -T_{ab}T^{ab} + T^2 + {1 \over 4 \lambda^2}\right). \label{genttbarcc}
\end{align}
where the generating functional is defined as $W = i \ln Z$ like before. For the same reasons outlined in the original \ttbar flow, the \ttbar part of the deformation to the generating functional differs by an overall sign compared to the deformation to the action. However, the cosmological constant term retains its sign.

Varying \eqref{genttbarcc} leads to the flow equation 
\begin{align}
    \partial_\lambda \left( \sqrt{\gamma} \delta \gamma^{ab}{{T}}_{ab} \right) = -2i \,  \delta \! \left( \sqrt{\gamma} \left( -{{T}}^{a b} {{T}}_{ab} + ({{T}}^a_a)^2 + {1 \over 4 \lambda^2} \right) \right) \label{variedflowCC}
\end{align}
where we used the Euclidean theory definition of the stress energy
\begin{align}
    \delta (- \ln Z) ={1 \over 2} \int \sqrt{\gamma} \,  T_{ab} \delta \gamma^{ab}.
\end{align}
This breaks up into the pair of equations
\begin{align}
    \partial_\lambda \gamma_{ab} = 4i \left( {{T}}_{ab} - \gamma_{ab} {{T}} \right) \equiv 4 i \hat{{{T}}}_{a b}, \quad \partial_\lambda \hat{{{T}}}_{ab} = 2 i \hat{{{T}}}_{ac}\hat{{{T}}}_{b}^{\ c} + i {\gamma_{ab} \over 4 \lambda^2}, \label{guicaflowCC}
\end{align}
but we no longer have $ \partial_\lambda (\hat{{{T}}}_{ac}\hat{{{T}}}_{b}^{\ c}) = 0$. We'd like to solve these equations subject to  continuity conditions when transitioning from the initial \ttbar flow. To analyze this, note that if we set $T_{ab} \rightarrow  -i \widetilde{T}_{ab}$, the flow equation \eqref{guicaflowCC} is identical to \eqref{guicaflow} up to the $\Lambda_2$ term, namely
\begin{align}
	    \partial_\lambda \gamma_{ab} =  4  \hat{{\widetilde{T}}}_{a b}, \quad \partial_\lambda \hat{{{T}}}_{ab} = 2  \hat{{\widetilde{T}}}_{ac}(\gamma^{-1})^{cd}\hat{{\widetilde{T}}}_{db} -{\gamma_{ab} \over 4 \lambda^2}. \label{guicaflowCCtilde}
\end{align}
Since the equations differ in the source term to $\partial_\lambda T \sim \partial_\lambda^2 \gamma$, we will impose continuity on the metric and its first derivative. 
The continuity condition on $\partial_\lambda T \sim \partial_\lambda^2 \gamma$ is subtle. Because $\mathrm{det} \gamma$ vanishes as noted around \eqref{detgammattbar}, the $ T \gamma^{-1} T$ term on the RHS is undetermined at the critical value $\lambda^-_c$: In order to fix it we need to introduce an extra condition regarding $\partial_\lambda T \sim \partial^2_\lambda \gamma$.\footnote{Compare with the exterior case where $\gamma^{-1}$ is well defined and one can observe from the equations of motion \eqref{guicaflow} that fixing $\gamma$ and $T \sim \partial_\lambda \gamma$ immediately fixes $\partial_\lambda T$ at $\lambda=0$.} The most naive condition is to set
\begin{align}
	\partial_\lambda^2 \hat{\widetilde{T}} = - {\gamma \over 4 \lambda^2} \delta (\lambda - \lambda_c),
\end{align}
but we checked that this does not reproduce the results of the operator method or those from the gravitational analysis in Sec.~\ref{sec:GPI}. Requiring this agreement leads to a very simple continuity condition, namely that $\partial_\lambda T \sim \partial_\lambda^2 \gamma$ be anti-continuous across $\lambda_c^-$, picking up an overall minus sign. To summarize, the boundary conditions are
\begin{align}
	\gamma |_{\lambda_c^<} =  \gamma |_{\lambda_c^>}, \quad \partial_\lambda \gamma |_{\lambda_c^<} = \partial_\lambda \gamma |_{\lambda_c^>}, \quad \partial_\lambda^2 \gamma |_{\lambda_c^<} = -\partial_\lambda^2 \gamma |_{\lambda_c^>} \label{outerhorizonbcs}
\end{align}
where $\lambda_c$ should be understood as  $\lambda_c^-$. Note that the middle condition is what imposes the continuity of the deformed energies.

Now we proceed to solve the flow equations \eqref{guicaflowCC}. By taking a derivative of the $\partial_\lambda T$ equation and plugging in the $\partial \gamma$ equation, we can transform these equations into a set of linear ordinary differential equations. Defining $\alpha = \ln[4 \lambda]$ and $g = e^{-\alpha}\gamma$ (understood as a matrix equation with indices suppressed), we find
\begin{align}
    \partial_\alpha g + g - i\hat{T} = 0, \quad \partial^2_\alpha \hat{T} - \partial_\alpha \hat{T} + 2 (\hat{T} + i g) = 0.
\end{align}
These equations can be solved using sines and cosines to give
\begin{align}
    \gamma &= 4 \lambda \hat{T}^{[0]} + 8 \hat{T}^{[0]} \eta^{-1} \hat{T}^{[0]} \lambda_c \lambda \cos \left[ \ln \lambda/\lambda_c\right] \label{flowed_metric_CC} \\
    \hat{T} &= -i \lambda \hat{T}^{[0]} -2 i \hat{T}^{[0]} \eta^{-1} \hat{T}^{[0]} \lambda_c \left(  \cos \left[ \ln \lambda/\lambda_c\right] -  \sin \left[ \ln \lambda/\lambda_c\right] \right)
\end{align}
where $\lambda_c$ is given in \eqref{lambdac} and the continuity conditions have already been implemented.

In order to reproduce the deformed energy of the operator method analysis, we need to transform the stress tensor to the appropriate canonical frame as discussed in section \ref{sec:review_gen_method}. A quick summary of the process is
\begin{enumerate}
    \item Find the coordinate transformation ${dY_\lambda \over dX}$ that maps the metric $\gamma$ to the flat Euclidean metric $\delta$ satisfying
    \begin{align}
         \left({dY_\lambda\over dX}\right)^T\cdot\delta \cdot\left({dY_\lambda\over dX}\right)  = \gamma.
    \end{align}
    This coordinate transformation relates the flowed stress tensor $T^{[\lambda]}$ to the physical one,
    \begin{align}
        {\cal T} = \left[ \left({dY_\lambda\over dX}\right)^T\right]^{-1} \cdot T^{[\lambda]} \cdot\left({dY_\lambda\over dX}\right)^T. \label{T_physicalT_cc}
    \end{align}
    This expresses the physical stress tensor as a function of $T^{[0]}$ and $\lambda$.
    \item Find the transformation $\Lambda$ that maps the canonical coordinates at $\lambda$ to those at $\lambda = 0$,
    \begin{align}
        dY_\lambda = \Lambda \cdot dY_0,
    \end{align}
    by imposing the two requirements: 1) that $\Lambda$ keeps the size of the compact direction fixed and 2) that the coordinate transformation between $X$ and $Y_0$ given by $\left({dY_\lambda\over dX}\right)^{-1} \cdot \Lambda$ is an element of $\mO(1,1)$.\footnote{It is $\mO(1,1)$ as opposed to $\mO(2)$ because the asymptotic boundary metric is Lorentzian.} The transformation between $X$ and $Y_0$ relates $T^{[0]}$ to the undeformed tensor in the coordinates $Y_0$ through
        \begin{align}
       T^{[0]} =  \left({dY_\lambda\over dX}\right)^T \cdot \Lambda^T \cdot {\cal T}^0 \cdot \Lambda \cdot \left[\left({dY_\lambda\over dX}\right)^T \right]^{-1} . \label{T0_physicalT0_CC}
    \end{align}
    \item Plug \eqref{T0_physicalT0_CC} into \eqref{T_physicalT_cc} to obtain ${\cal T}$ as a function of ${\cal T}^0$ and $\lambda$.
\end{enumerate}
Following these steps, we obtain
\begin{align}
    {\cal T}_{tt} &= {i \over 4 \lambda} \left( 1 - \sqrt{-1 + 8 \lambda {\cal T}^0_{tt} - 16 \lambda^2 ({\cal T}^0_{tx})^2}\right), \\
    {\cal T}_{tx} &= -i{\cal T}_{tx}^0, \\
    {\cal T}_{xx} &={i \over 4 \lambda} { 1 - 16 \lambda^2 ({\cal T}^0_{tx})^2+\sqrt{-1 + 8 \lambda {\cal T}^0_{tt} - 16 \lambda^2 ({\cal T}^0_{tx})^2}  \over \sqrt{-1 + 8 \lambda {\cal T}^0_{tt} - 16 \lambda^2 ({\cal T}^0_{tx})^2}},
\end{align}
providing agreement with the results of the previous section.

We conclude by analyzing the signature of the flowed metric. By directly computing the determinant of the expression in \eqref{flowed_metric_CC} and expressing it in terms of the undeformed stress tensor ${\cal T}^0$, we find
\begin{align}
    \mathrm{det}[\gamma] = N^2 \left[ -1 + 8 \lambda {\cal T}^0_{tt} - 16 \lambda^2 ({\cal T}^0_{tx})^2\right],
\end{align}
for some positive $N^2$. Therefore the determinant is positive within the range $\lambda_c^- < \lambda < \lambda_c^+$, indicating that its signature has changed from the region $\lambda < \lambda_c^-$, and vanishes at $\lambda_c^\pm$ as expected.

\subsection{Beyond $\lambda_c^+$}

Continuing the \ttbar + $\Lambda_2$ flow beyong $\lambda_c^+$ generates complex values of the deformed energy. This can be avoided by reverting back to the original \ttbar deformation at $\lambda_c^+$, namely
\begin{align}
    \partial_\lambda S_\lambda&=   \int \! \! dt d\theta \left( T_{ab}T^{ab} - T^2 \right),
\end{align}
and imposing similar continuity conditions as those at $\lambda_c^-$. As a result, the signature switches back to being Lorentizian and the analysis is identical to that of section \ref{sec:review}. To quote the results, the deformed energy will again be given by
\begin{align}
	E_\lambda = {L \over 4 \lambda} \left( 1 - \sqrt{1 - 8 \lambda E_0/L + 64 \pi^2 J^2 \lambda^2/L^4} \right).
\end{align}
Notice that this remains real for all $\lambda > \lambda_c^+$. We see that for all $E_0$, there is an accumulation point where all deformed energies tend to $-2 \pi J/L $ as $\lambda \rightarrow \infty$. The determinant of the metric in the generating functional approach given by
\begin{align}
    \mathrm{det}[\gamma]  = -\left( {2 \sqrt{1 - 8 \lambda {\cal T}^0_{tt} + 16 \lambda^2 ({\cal T}^0_{tx})^2} \over 1- 4{\cal T}^0_{tx} + \sqrt{1 - 8 \lambda {\cal T}^0_{tt} + 16 \lambda^2 ({\cal T}^0_{tx})^2}} \right)^2
\end{align}
exhibits  similar behavior tending to $-4$ without changing its signature. In what follows we show that this is signals the approach to the $r = 0$ singularity in the BTZ black hole.

\begin{figure}[ht]
    \centering
    \includegraphics[width=0.6\linewidth]{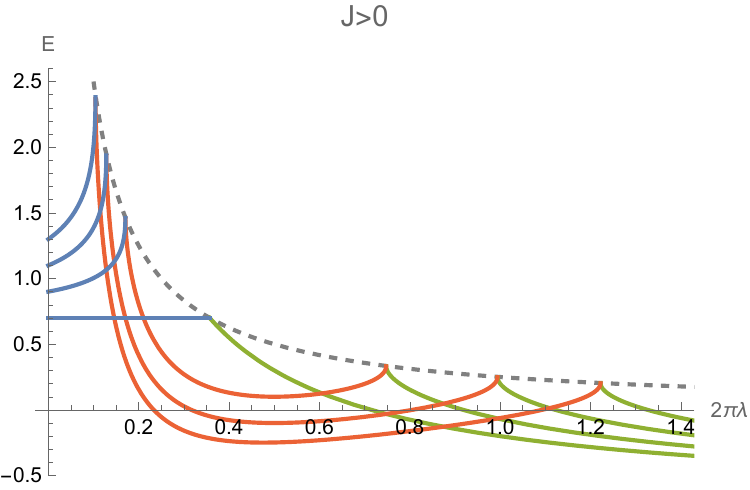}
    \caption{The extended flowed energy levels for various different seed energies $E_0$ across $\lambda^+_c$. We only show the $J>0$ plot here since $\lambda^+_c=\infty$ when $J=0$. The extension of the flowed energy levels beyond $\lambda^+_c$ are shown in green. It is worth noting that the extremal case (where $E_0=J$, as indicated by the fourth curve from the top) does not experience the second flow (red part) since $\lambda^-_c=\lambda^+_c$ for the extremal states.}
    \label{fig:energy_level_deep}
\end{figure}

\section{Boundary deformations from bulk flows}
\label{sec:bulkflow}

 We will now use the bulk to motivate the sequence of deformations analyzed in the previous two sections. We will review how to show that the original \ttbar flow reviewed in section \ref{sec:review} corresponds to moving the holographic boundary inwards from the asymptotic boundary, and then determine what kind of bulk operation is dual to the \ttbar + $\Lambda_2$ flow analyzed in section \ref{sec:ttbarCC}. A main takeaway will be that the presence of the $\Lambda_2$ term correlates with the signature of the normal to the holographic boundary.

Following the techniques from \cite{Hartman:2018tkw,Taylor:2018xcy}, we will consider a bulk foliation by codimension one surfaces that corresponds to moving the holographic boundary throughout the bulk, and use the Hamiltonian constraint generating this flow to derive the boundary interpretation. The Hamiltonian constraint on any codimension one surface $\Sigma$ is given by $G_{ab}n^an^b$ where $G_{ab}$ is the Einstein tensor and $n^a$ is the unit-normal to the surface pointing in the direction of the flow. For Einstein-Hilbert with negative cosmological constant, it reads
\begin{align}
	{1 \over 2} \left( R_{\Sigma} + K^2 - K_{ab}K^{ab} - 2 n_an^a \right) = 0
\end{align}
The term proportional to $n_an^a$ comes from the bulk cosmological constant which we set to one.  This term is either $\pm1$ depending on the signature of the normal. The extrinsic curvature is defined relative to $-n^a$ which points outward, opposite from the direction of flow. Finally, $R_\Sigma$ is the intrinsic Ricci scalar on $\Sigma$.

The procedure for deriving the flow is to express the Hamiltonian constraint in terms of the Brown-York (BY) stress tensor computed at the boundary of the spacetime, and then find the kind of boundary flow to which it corresponds. We will restrict our analysis here to three dimensions, although the generalization is straightforward.
\subsection{Recap: boundary with a spacelike normal}
The BY stress tensor is defined by varying the bulk action with respect to the induced metric on the boundary. We restrict to the following Lorentzian action with a timelike boundary
\begin{align}
    Z = \int {\cal D}\psi \,  e^{-i S_B} \rightarrow S_{B} = - {1 \over 2}\int \sqrt{-g} \left( R - 2 \right) - \int \sqrt{- h} (K - 1). \label{action_exterior}
\end{align}
The minus sign in the boundary term is a counterterm given by the (negative) proper area of the boundary. We set $8 \pi G = 1$. The BY tensor is
\begin{align}
    \widetilde{T}_{ab} &= i {2 \over \sqrt{-h}}  {\delta \over \delta h^{ab}} \ln Z  \label{tlnzlorentzian} \\
    &= {2 \over \sqrt{-h}} {\delta \over \delta h^{ab}}S_B|_\mathrm{on-shell}, \\
    &= -K_{ab} + h_{ab} \left( K - 1 \right).
\end{align}

This formula can be used to replace the extrinsic curvatures appearing in the constraint equation with the BY tensor. This gives the ``trace flow" equation  expressing the trace of the BY tensor $\widetilde{T}\equiv \widetilde{T}^a_a$ in terms of the square of the stress tensor
\begin{align}
     \widetilde{T} = -{1 \over 2}\left(R_\Sigma -\widetilde{T}_{ab}\widetilde{T}^{ab} + \widetilde{T}^2 \right).
\end{align}All the tensors in this equation are bulk quantities. To translate this equation into a boundary statement, we renormalize the BY stress tensor and the induced metric by their boundary duals. We will focus on holographic boundaries with vanishing intrinsic Ricci scalar $R_\Sigma$. Assuming that the bulk metric takes the Fefferman-Graham (FG) form
\begin{align}
    ds^2 = {d \rho^2 \over 4 \rho^2} + {\gamma_{ab} \over \rho}dx^a dx^b,
\end{align}
where $\gamma_{ab}$ is the physical boundary metric, then the renormalization prescription (special to three dimensions) is
\begin{gather}
    h_{ab} = {\gamma_{ab} \over \rho}, \quad  \widetilde{T}_{ab} =  T_{ab}, \quad \widetilde{T} = \rho T, \quad \widetilde{T}^{ab} = \rho^2 T^{ab}.
\end{gather}
This leads to the boundary trace flow equation 
\begin{align}
    T = -{ \rho \over 2 }\left( -T_{ab}T^{ab} + T^2 \right).
\end{align}
This equation is not a purely boundary object yet because it depends on the bulk radial parameter $\rho$. We will argue that $\rho$ should be understood as the boundary deformation parameter. To see this, note that since there is only one scale in the boundary theory, increasing the deformation parameter from $\lambda$ to $\lambda + \delta \lambda$ amounts to an infinitesimal scale transformation $x^\mu \rightarrow (1 - \delta \lambda/2\lambda) x^\mu$. From the definition of the stress tensor \eqref{tlnzlorentzian}, we must have
\begin{align}
    \partial_\lambda (i\ln Z) = {1 \over 2 \lambda} \int \! \! \sqrt{-\gamma} \ T,
\end{align}
or in terms of the actionF
\begin{align}
    \partial_\lambda S_\mathrm{QFT} &= {1 \over 2 \lambda} \int \! \! \sqrt{-\gamma} \ T.
\end{align}
If we take $\rho = 4 \lambda$ we end up with
 \begin{align}
    \partial_\lambda S_\mathrm{QFT} &= \int \! \! \sqrt{-\gamma} \left( T_{ab}T^{ab} - T^2 \right),
\end{align}
reproducing the standard \ttbar deformation.

\subsubsection{Application to the BTZ exterior and extracting the deformed energy}
Next, we review how to obtain the \ttbar deformed energy formula from the BY tensor on the BTZ background. The renormalized boundary metric in FG gauge is
\begin{equation}
    \gamma_{ab} = \frac{1}{4}\begin{pmatrix}
        -r_+^2(\rho-1)^2+r_-^2(\rho+1)^2 & 4r_+r_-\rho \\
        4r_+r_-\rho &r_+^2(\rho+1)^2-r_-^2(\rho-1)^2
    \end{pmatrix}
\end{equation}
One can check that $\gamma_{ab}$ solves the metric flow equation \eqref{guicaflow} with the renormalized (trace-reversed) stress tensor
\begin{equation}
    \hat{T}_{ab} \equiv T_{ab}-\gamma_{ab}T = \frac{1}{2}
    \begin{pmatrix}
        -r_+^2(\rho-1)+r_-^2(\rho+1) & 2r_+r_-\rho \\
        2r_+r_-\rho &r_+^2(\rho+1)-r_-^2(\rho-1)
    \end{pmatrix}
\end{equation} 
with the identification $\rho=4\lambda$.

To extract the deformed energy, we need to pass to the physical boundary metric in BTZ coordinates. The BTZ radial coordinate $r$ is related to the FG radial $\rho$ through
\begin{align}
    {dr^2 \over f(r)} = {d \rho^2 \over 4 \rho^2}.
\end{align}
Note that the overall normalization of $\rho$ is not fixed by this relation. It can be determined by requiring that the size of the boundary circle parameterized by $\theta$ remains fixed along the flow. Hence, at the cutoff surface, we need $\rho_c = 1/r^2_c$. The solution is
\begin{align}
    \rho = {r_c^{-2}} {2 r^2 - r_+^2 - r_-^2 - 2 r \sqrt{f(r)} \over 2 r_c^2 - r_+^2 - r_-^2 - 2 r_c \sqrt{f(r_c)}}.
\end{align}
Therefore, the physical boundary metric is
\begin{align}
    -{f(r_c) \over r_c^2}dt^2 + \left(d\theta - {r_+ r_- \over r_c^2} dt \right)^2. \label{btzboundarymetric}
\end{align}

The components $T_{ab}$ of the boundary stress tensor in the previous subsection are those in the frame set by the metric \eqref{btzboundarymetric}. To compare with the boundary, we need to work in the ``canonical'' frame where the metric is $-d\tilde{t}^2 + d\tilde{\theta}^2$. Therefore, we need to transform the stress tensor to this frame. This is done through the coordinate transformation
\begin{align}
    \widetilde{t} = \sqrt{f(r_c) \over r_c^2} t, \quad \tilde{\theta} = \theta - {{r_+ r_- \over r_c^2} }t.
\end{align}
The stress tensor in this canonical frame is given by
\begin{align}
    \tilde{T} = \left( {d {x} \over d \tilde{x}} \right)^\mathrm{T} \cdot T \cdot {d {x} \over d \tilde{x}}.
\end{align}
The energy and angular momentum can be read off from the components of the canonical stress tensor to be
\begin{align}
    E &= 2 \pi \widetilde{T}_{\tilde{t} \tilde{t}} = 2 \pi  r^2 \left( 1 - \sqrt{f(r)/r^2}\right)= {2 \pi \over 4 \lambda} \left( 1 - \sqrt{1 -  {8 \lambda M \over 2 \pi} +  {16 \lambda^2 J^2 \over (2 \pi)^2}}\right),  \\
    J &= -{2 \pi \widetilde{T}_{\tilde{t} \tilde{\theta}}}  = 2 \pi r_+ r_-.
\end{align}
where $M = 2 \pi (r_+^2 + r_-^2)/2$. This reproduces precisely the form of the energy and angular momentum in a \ttbar deformed CFT. 
\subsection{Boundary with timelike normal}
We extend the analysis of the previous section to situations where the boundary has a timelike normal. This will be relevant for boundaries placed at ``radial'' slicing in the BTZ interior. The bulk action we  consider is
\begin{align}
    S_{B} = - {1 \over 2}\int \sqrt{-g} \left( R - 2 \right) + \int \sqrt{ h} (K - 1),
\end{align}
where we flip the sign of the boundary term from the usual case to account for Stokes' theorem  for boundaries with a timelike normal. Since the normal is timelike, the boundary signature is spacelike. This change of signature and sign of the boundary term should be interpreted as the bulk analogue of the sign change in the time evolution operator discussed in section \ref{sec:ttbarCC_Op}.

Since the holographic boundary is purely spacelike, we will treat the boundary theory as a Euclidean theory. Hence the BY tensor is defined as
\begin{align}
    \widetilde{T}_{ab} &= - {2 \over \sqrt{h}} {\delta \over \delta h^{ab}} \ln Z \\
    &= i {2 \over \sqrt{h}} {\delta \over \delta h^{ab}}S_B|_\mathrm{on-shell}, \\
    &=i \left(  K_{ab} - h_{ab} (K - 1)\right).
\end{align}
The stress tensor picks up an overall $i$ because, while the boundary metric is Euclidean, there's an extra $i$ multiplying the action. 

As before, we derive a trace flow equation by re-expressing the extrinsic curvatures in the Hamiltonian constraint in terms of the BY tensor. Note that the constant term in the Hamiltonian constraint now comes with a different sign since $n^2 = -1$. As a result, the expression of the Hamiltonian constraint in terms of the BY tensor gains an additional term
\begin{align}
     -i\widetilde{T} = {1 \over 2}\left(R_\Sigma + \widetilde{T}_{ab}\widetilde{T}^{ab} - \widetilde{T}^2  + 4\right).
\end{align}
To obtain this from the boundary, we consider the bulk in FG form
\begin{align}
        ds^2 = -{d \rho^2 \over 4 \rho^2} + {\gamma_{ab} \over \rho}dx^a dx^b,
\end{align}
and take the surfaces $\Sigma$ to have vanishing intrinsic scalar curvature. Then the trace flow equation renormalizes to
\begin{align}
    -iT &= { \rho \over 2 }\left( T_{ab}T^{ab} - T^2 + {4 \over \rho^2}\right),\\
    &={ 4 \lambda \over 2 }\left( T_{ab}T^{ab} - T^2 + {1 \over 4 \lambda^2}\right),
\end{align}
where in the second line we use the relation $\rho = 4 \lambda$.

As before, the trace of the stress tensor can be related to the derivative of the partition function. In a Euclidean theory, this is 
\begin{align}
 \partial_\lambda (-\ln Z) = {1 \over2 \lambda} \int \sqrt{\gamma} \, T.
\end{align}
Using the trace flow equation on the right hand side and expressing the action in terms of the action on the left we get
 \begin{align}
    \partial_\lambda S_\mathrm{QFT} &= \int \! \! \sqrt{\gamma} \left( T_{ab}T^{ab} - T^2 + {1 \over 4 \lambda^2}\right).
\end{align}
This differs from the standard \ttbar expression by a  $\lambda$-dependent boundary $\Lambda_2$ term and is precisely the deformation analyzed in section \ref{sec:ttbarCC}.

\subsubsection{The deformed energy between the inner and outer horizons}

In the interior region, the renormalized boundary metric in the FG gauge reads 
\begin{equation}
    \gamma_{ab} = \frac{\rho}{2}
    \begin{pmatrix}
    r_+^2 + r_-^2 - (r_+^2-r_-^2)\cos(\ln \rho) & 2r_+r_- \\
    2r_+r_- & r_+^2 + r_-^2 + (r_+^2-r_-^2)\cos(\ln \rho)
    \end{pmatrix}
\end{equation}
One may be puzzled here that $\gamma_{ab}$ is not quadratic in $\rho$. This is because that the relation between FG and BTZ radial coordinate picks up a minus sign in the interior, as in
\begin{equation}
    \frac{dr^2}{f(r)} = -\frac{d\rho^2}{4\rho^2}, \quad\quad r_-<r<r_+.
\end{equation}
Similar to the exterior case, one can check that, together with the boundary stress tensor $T_{ab}$, $\gamma_{ab}$ satisfies the metric flow equation \eqref{guicaflowCC} with the $\Lambda_2$ term after identifying $\rho=4\lambda$.

We now find the stress energy tensor in the ``canonical'' frame for the boundary metric between the inner and outer horizons. Since constant BTZ radial slices are purely spacelike, the canonical metric is $+d\tilde{t}^2 + d\tilde{\theta}^2$.  The coordinate transformation is now
\begin{align}
    \widetilde{t} = \sqrt{-f(r_c) \over r_c^2} t, \quad \tilde{\theta} = \theta - {{r_+ r_- \over r_c^2} }t.
\end{align}


We get the stress tensor in the canonical frame by applying this coordinate transformation. The energy is given by the generator of time evolution, given here as $-2 \pi i \widetilde{T}_{\tilde{t} \tilde{t}}$. The angular momentum is the same as it would be in a Euclidean theory, namely $-{2 \pi i \widetilde{T}_{\tilde{t} \tilde{\theta}}}$.  We find 
\begin{align}
    E &= -2 \pi i \widetilde{T}_{\tilde{t} \tilde{t}} = 2 \pi  r^2 \left( 1 - \sqrt{-f(r)/r^2}\right)= {2 \pi \over 4 \lambda} \left( 1 - \sqrt{-1 +  {8 \lambda M \over 2 \pi} -  {16 \lambda^2 J^2 \over (2 \pi)^2}}\right) \\
    J &= -{2 \pi i \widetilde{T}_{\tilde{t} \tilde{\theta}}}  = 2 \pi r_+ r_-.
\end{align}
These are consistent with the deformed energy and angular momentum found using the boundary deformation with the $\Lambda_2$ analyzed in Sec.~\ref{sec:ttbarCC}. Note that both the angular momentum and energy are continuous at the outer horizon where $f(r) = 0$.



\subsection{Between the inner horizon and the singularity}

In the region behind the inner horizon, the constant radial slices are timelike again, and the analysis of the exterior case carries through to the region behind the inner horizon. Therefore, the deformation that flows from the inner horizon to the singularity is the standard \ttbar deformation without the boundary $\Lambda_2$ term, but with initial conditions at the inner horizon set by the previous flow between the inner and outer horizons.

\subsection{Summary}

We used the bulk to motivate the following prescription for deforming the holographic CFT to explore all regions of the BTZ black hole. If we define the stress tensor as
\begin{align}
    T_{ab} &= {2 \over \sqrt{|\gamma|}} {\delta \over \delta \gamma^{ab}}S_B|_\mathrm{on-shell}
\end{align}
throughout the evolution (i.e. we take $T \rightarrow  i T$ for the timelike normal boundary) then the sequence of deformations is
 \begin{align}
    \partial_\lambda S_\mathrm{QFT} &= \int \! \! \sqrt{-\gamma} \left( T_{ab}T^{ab} - T^2 \right), \quad \  \quad \quad \ \ \ &&0 < \lambda < {1 \over r_+},\\
    &= \int \!\! \sqrt{\gamma} \left( -T_{ab}T^{ab} + T^2 + {1 \over 4 \lambda^2}\right),  \quad &&{1 \over r_+}  < \lambda < {1 \over r_-}, \\
    &= \int \! \! \sqrt{-\gamma} \left( T_{ab}T^{ab} - T^2 \right), \quad \quad \quad \  \quad &&{1 \over r_-} < \lambda < \infty.
\end{align}

\section{On the bulk implementation of \ttbar}
\label{sec:GPI}
In this section, we shift to a purely bulk perspective and study the bulk interpretation of the flows described in the previous sections. The goal is to define a gravitational path integral along with the appropriate boundary conditions to give the bulk dual of a \ttbar deformed theory in all the regimes discussed in the previous sections. Instead of imposing Dirichlet boundary conditions on the entire induced metric at the finite cutoff surface, as is usually done in this subject, we will impose mixed boundary conditions that fix the quasi-local and asymptotic gravitational charges consistent with the boundary theory. We will show that these boundary conditions give the correct saddles for both the exterior and interior of the black hole. We will also discuss the purely Dirichlet problem and use it to motivate a new definition for the canonical ensemble of a deformed boundary theory.
We start with a generic discussion for gravity in generic dimensions in Sec~\ref{sec:GPI-gen} before specializing to the 3d  case in Sec~\ref{sec:GPI-BTZ} and the JT gravity case in Sec~\ref{sec:GPI-JT}. We define and analyze the deformed canonical ensemble in Sec \ref{sec:GPI-can}.
\subsection{Generic construction}
\label{sec:GPI-gen}

The boundary protocol for implementing \ttbar begins with a particular microstate labelled by an energy $E_0$ and angular momentum $J_0$.
The flow equation for the \ttbar deformation then describes for us how the energy changes as we switch on the deformation. This  uniquely determines a new flowed energy $E(\lambda)$ by the requirement of translation invariance (which corresponds to spherical symmetry in the bulk) given the following three parameters:
\begin{enumerate}[(1)]
    \item The undeformed energy $E_0$.
    \item The undeformed angular momentum $J_0$.
    \item The deformation parameter $\lambda$.
\end{enumerate}
In the bulk, these will map to (1) ADM mass $M$, (2) angular momentum $J$, (3) a function of the spatial component of the induced metric. With those quantities fixed, the gravitational path integral we define will amount to specifying a corresponding cutoff bulk with fixed quasi-local energy on the boundary.\footnote{This should be thought of as a microcanonical ensemble with a ${\cal O}(G_N^0)$  width.}

Let us outline our setup. We will analyze the problem in  arbitrary dimensions before specializing to three dimensions. As usual, we consider the Einstein-Hilbert action
\begin{equation}
S = -\frac{1}{16\pi G}\int_\mathcal{M}d^{d+1}x\sqrt{g}(R+2) -  \frac{s}{8\pi G}\int_{\partial\mathcal{M}}d^dx \sqrt{|\gamma|}(K-1)+ S_\mathrm{micro},
\end{equation}
where the manifold $\mathcal{M}$ has boundary $\partial \mathcal{M}$ and where $s = +1/-1$ when the signature of the normal is spacelike/timelike.\footnote{The sign is from the dependence  Stokes' law on the signature of the boundary normal.}
The additional boundary term $S_\mathrm{micro}$ will be chosen to set up a well-posed variational problem with the desired set of boundary conditions. Let $\gamma_{ab}$ be the induced metric on the boundary and $\gamma$ be its determinant. Consider a foliation of the boundary by Cauchy surfaces $\Sigma_t$ labelled by a time parameter $t$ and define the time flow vector field $t^a = \nabla^a t$.\footnote{This is a spacelike vector if the boundary metric is Euclidean. This can be continued  $t^a \rightarrow i t^a$ for the Lorentzian case.} Let $n^a$ be the normal to $\Sigma_t$ and $\sigma_{ab} \equiv \gamma_{ab} - n_a n_b$ be the induced metric on $\Sigma_t$.

One set of boundary conditions is to fix the intrinsic geometry of $\Sigma_t$. This intrinsic geometry can be described by the induced metric in Gaussian normal coordinates, for example, which we can denote by  $\tilde{\sigma}_{mn}$.\footnote{Since normal geodesics to a Cauchy slice might eventually intersect, these coordinates have to be defined separately for each $\Sigma_t$.} In the holographic context, the metric of the dual theory is related to the bulk metric by an over all scalar renormalization factor  $\Omega$ and where  $\tilde{\sigma}_{mn} = \Omega \, \hat{\sigma}_{mn}$ where $\hat{\sigma}_{mn}$ is the physical boundary metric. While $\Omega$ is commonly taken to infinity, we will assign it a  finite value as part of our boundary conditions. The finiteness of $\Omega$ is what is referred to as ``finite cutoff holography.'' It will be inversely related to the \ttbar deformation parameter.

The next set of boundary conditions on the finite cutoff surface fixes the asymptotic ADM charges.  While this seems like a contradiction in terms, it is achievable for spherically symmetric metrics due to a Gauss law imposed by the gravitational constraints relating the asymptotic charges to local data anywhere in the spacetime \cite{Kenmoku:2001qt,Kenmoku:2002nu,Kuchar:1994zk}. The ADM charges can be written as locally conserved quantities  in terms of local metric data.
\begin{gather}
	M = {\Omega \over 8 G} \left( \sigma^{ac}(K_{ab}n^b)(K_{cd}n^d) - s (K_{ab}\sigma^{ab})^2 + 1 \right), \\
	J_c = {\Omega \over 4 G}  \,K_{ab}n^a \sigma^b_{\ c},
\end{gather}
where these extrinsic curvatures describe the embedding of the boundary surface in the bulk. These ADM charges are ``locally conserved'' in the sense that their derivative in the normal direction to the boundary vanishes, something which follows from the Hamiltonian and momentum constraints \cite{Kuchar:1994zk}.  The expression for $M$ is related to a quantity called the Hawking mass, see \cite{Folkestad:2022dse, Folkestad:2023cze,Soni:2024aop} for a discussion on this and an application to \ttbar. 

These charges are somewhat complicated functionals of the boundary data and are  tricky to impose directly in the path integral. A more standard procedure  is to fix the components of the extrinsic curvature appearing in the ADM charges, namely $K_{ab}n^b \sigma^{a}_c$ and $K_{ab}\sigma^{ab}$. Note that this is the same as fixing the quasi-local Brown-York charges
\begin{gather}
	E/2 \pi =  -{ \Omega \over \sqrt{s}} \,   T_{ab} n^a n^b = {\Omega \over 8 \pi G} \, (K_{ab} - h_{ab}(K - 1))n^{a} n^{b} = {\Omega \over 8 \pi G} \left( -K_{ab}\sigma^{ab} + 1 \right),  \\
	  J_c/2 \pi = -{\Omega \over \sqrt{s} } \, T_{ab}n^a \sigma^b_{\ c} =  {\Omega \over 8 \pi G} \,(K_{ab} - h_{ab}(K - 1))n^a \sigma^b_{\ c} = {\Omega \over 8 \pi G} \,K_{ab}n^a \sigma^b_{\ c}.
\end{gather}
These expressions allow us to relate all the boundary data we want to fix with a single equation stating
\begin{equation}
\label{eq:MJEh}
M = \frac{1}{8G}\left(\frac{16G^2J^2}{\Omega}- \frac{(4GE-\Omega)^2}{\Omega}+\Omega\right).
\end{equation}
This equation shows that we can fix three $(M,J,E,\Omega)$ and solve to find the fourth. However, there could be multiple solutions. For example, if we fix $(M,J,\Omega)$ then if $E$ is a solution  so is $\Omega/2 - E$. These two solutions correspond to changing the over all sign of the extrinsic curvature, which can be thought of assigning which side of the finite cutoff defines the bulk. If we solve for $E$ for the two cases we get the \ttbar flowed energies
\begin{align}
	E = {\Omega \over 4 G}\left( 1 \pm \sqrt{s \left[ 1 - 8G M/\Omega  +   16 G^2 J^2/\Omega^2\right]}\right).
\end{align}
This says two things: that we should identify $G/\Omega$ with the deformation parameter $\lambda$, and that if we define the path integral with fixed $(M,J,h)$ then we must sum over both branches of the square root. 

Next, we need to pick the appropriate $S_\mathrm{micro}$ that sets up a well-posed variational problem given our chosen boundary conditions. Since we are fixing the energy and angular momentum, our path integral defines a micro-canonical ensemble. This ensemble is just the Laplace transform of the canonical ensemble, and so we need to include the analogue of $\beta (E + \omega J)$ into the action and integrate over $\beta$ and $\omega$. We can achieve this by including in the boundary action the term
\begin{align}
	S_\mathrm{micro}  =   \frac{s}{8\pi G}\int_{\partial\mathcal{M}}\sqrt{|\gamma|} \, \, \, {T_{ab}n^a t^b \over n\cdot t}=  \frac{s}{8\pi G}\int_{\partial\mathcal{M}}{\sqrt{|\gamma|} } \,  \left( {T_{ab}n^a n^b + T_{ab}n^a {N^b \over n\cdot t} }    \right),
\end{align}
where $N^b$ is the shift. The factor $n\cdot t$ is equal to the lapse and the ratio $N^a/n\cdot t$ appropriately normalizes the rotational or translational speed of the time flow. 

Finally, we can put all this together to define the micro-canonical finite cutoff gravitational path integral that fixes $(M,J,\Omega)$. We found  four options consistent with this data: the choice of  $s = \pm 1$ representing the signature of the boundary normal, and the choice of branch of the square root in the  quasi-local energy. A natural proposal for the path integral is to sum over all these options as
\begin{align}
	Z_\mathrm{All} \equiv \int \! {\cal D}g \!\sum_{s,m = \pm} e^{-S_\mathrm{EH} - {s }  \left(  S_\mathrm{BY} - S_\mathrm{micro}^{s,m} \right)},
\end{align}
where $m = \pm$ keeps track of the choice of branch in the quasi-local energy, and where the extra boundary term is
\begin{align}
	S_\mathrm{micro}^{s,m} = \frac{1}{8\pi G}\int_{\partial\mathcal{M}}\sqrt{|\gamma|} \, \, \, {\widetilde{T}_{ab}^{s,m} n^a t^b \over n\cdot t},
\end{align}
where $\widetilde{T}_{ab}^{s,\pm}$ are the stress tensor components of the two branches. However, note that depending on the boundary conditions, not all choices for $s$ and $m$ admit real classical solutions. For instance, for $s = m = +1$ and sufficiently large $ \Omega$, the only solution to Einstein's Equations that admits the resulting Brown-York charges will be those on a finite cutoff surface where the normal points away from the asymptotic boundary. But such solutions with a second boundary are disallowed since the action contains only a single boundary term. Perhaps such an extra boundary term should be introduced by hand, but then this raises the question of whether this introduces a second boundary holographic system. We will exclude this possibility from our analysis. We will mostly avoid this issue altogether by restricting to the minus branch ($m = -1$) which always seems to have a solution. We do so by restricting the definition of the partition function to
\begin{align}
    Z_{-} \equiv \int \! {\cal D}g \!\sum_{s = \pm} e^{-S_\mathrm{EH} - {s }  \left(  S_\mathrm{BY} - S_\mathrm{micro}^{s,-1} \right)}. \label{Zminus}
\end{align}

Note that for generic real values of $(M,J,\Omega)$, the quasi-local energy will be complex for one of the signs of $s$ since it selects the sign of the argument of the square root. Therefore, if the path integral only integrates over spacetimes with real values of the quasi-local energy, then the Laplace transform will automatically set to zero the contribution with complex boundary term. Therefore, $s$ should be determined by $(M,J,\Omega)$ and the gravity path integral automatically implements the multi-step flow discussed in the previous sections.


\subsection{Finite cutoff in three dimensions}
\label{sec:GPI-BTZ}
Here we apply the above framework to  three dimensions. We present explicit expressions for the boundary terms, boundary conditions, and the types of solutions they generate. We take our metric ansatz to be the following generic stationary 3d  bulk with spherical symmetry
\begin{equation}
\label{eq:metricansatz}
ds^2 = f(r)d\tau^2 + g(r)dr^2 + h(r)(d\theta+N_{\theta}(r)d\tau)^2,
\end{equation}
where we chose to gauge-fix the metric such that the radial normal $\partial_r$ is perpendicular to both $\partial_\theta$ and $\partial_\tau$\footnote{This choice is always possible at least locally near $r=r_c$, which is sufficient for our purpose here.}. The boundary $\partial \mathcal{M}$ will be placed at constant $r$. The ADM charges in this parametrization \eqref{eq:metricansatz}  read as 
\begin{align}
\label{eq:ADMansatz}
M &= \frac{1}{8G} \left( -\frac{h(r)^2N'_\theta(r)^2}{4f(r)g(r)} - \frac{h'(r)^2}{4g(r)h(r)}+h(r) \right), \\
\label{eq:Jansatz}
iJ &= \frac{1}{8G}\frac{h(r)^{3/2}N'_\theta(r)}{\sqrt{f(r)g(r)}}.
\end{align}
We demonstrate that these charges satisfy $\partial_r M = \partial_r J=0$ in appendix \ref{App:conservation}.

Following the same procedure as in Sec.~\ref{sec:bulkflow}, we want to compute the BY stress tensor at constant $r$ in the canonical boundary frame. The relation between the canonical frame and the induced metric is
\begin{align}
    d\tilde{\tau}^2 + d\tilde{\theta}^2 = {f \over h}d\tau^2 + \left( d\theta + N_\theta d\tau\right)^2, \quad \tilde{\tau} \equiv \sqrt{f \over h} \,\tau, \quad \tilde{\theta} \equiv \theta + N_\theta \, \tau.
\end{align}
The BY stress tensor in the $\tilde{\tau}, \tilde{\theta}$ coordinates reads
\begin{align}
\label{eq:BYansatz}
 \widetilde{T}_{BY} = -\frac{\sqrt{s}}{8\pi G}
 \begin{pmatrix}
 h(r) - \frac{h'(r)}{2\sqrt{|g(r)|}} & \frac{h(r)^{3/2}N'_\theta(r)}{2\sqrt{|f(r)g(r)|}} \\
 \frac{h(r)^{3/2}N'_\theta(r)}{2\sqrt{|f(r)g(r)|}} & h(r) - \frac{f'(r)h(r)}{2f(r)\sqrt{|g(r)|}}
 \end{pmatrix}.
\end{align}
We can now identify
\begin{equation}
\label{eq:Eansatz}
E = -\frac{2\pi}{\sqrt{s}}  \widetilde{T}_{\tilde{\tau}\tilde{\tau}} = \frac{1}{4G}  \left(h(r) - \frac{h'(r)}{2\sqrt{|g(r)|}}\right).
\end{equation}
Note that the off-diagonal terms in \eqref{eq:BYansatz} are equal to $J/2\pi$ (up to a possible $i$), as expected. The quasi-local energy and angular momentum are fixed by the boundary term
\begin{equation}
    S_\mathrm{micro} = \frac{1}{2\pi}\int d^2x \sqrt{|\gamma|}\left( \frac{E}{h(r)}+i 
    \frac{N_\theta J}{\sqrt{|f(r)h(r)|}}\right)
\end{equation}
We check that this choice of boundary term indeed gives rise to a well-defined variation problem in Appendix~\ref{App:variation}.

Next we analyze the solutions for different ranges of $(M,J,h)$ where we replaced $\Omega$ with $h$. We restrict to purely gravitational solutions allowing only for black holes and conical defects, all of which are described by the static and spherically symmetric solution
\begin{equation}
ds^2 = f(r)d\tau^2 + \frac{dr^2}{f(r)} + r^2\left(d\theta-\frac{4iGJ}{r^2}d\tau\right)^2,\quad 0\le r\le \sqrt{h},
\end{equation}
with
\begin{equation}
f(r) = r^2-8GM+\frac{16G^2J^2}{r^2} = {(r^2 - r_+^2)(r^2 - r_-^2) \over r^2},
\end{equation}
where
\begin{align}
	M = {r_+^2 + r_-^2 \over 8 G}, \quad J = {r_+ r_- \over 4 G}.
\end{align}
and where $\sqrt{h}$ plays the role of the cutoff radius $r_c$.
\begin{center}
{ \bf Conical defects}
\end{center}
We start with the case of conical defects which lie in the mass range $-{1 \over 8 G} \le  M < 0$, $|J| \le M$. The deformed energy
\begin{align}
		E = {h \over 4 G}\left( 1 - \sqrt{ 1 + 8G |M|/h  +   16 G^2 J^2/h^2}\right)
\end{align}
is real for all values of $h$ since the argument of the square root remains positive (given that $s = +1$). The limit of $h \rightarrow 0$ brings the boundary all the way down to the conical defect. All energies approach the accumulation point $-J$ as the boundary approaches the conical singularity. We will see that similar behavior follows when approaching the black hole singularity.

\begin{center}
{ \bf Rotating non-extremal BTZ}
\end{center}
For the range $0 <  M < \infty$, $0<|J| < M$, the bulk solution is given by some patch of the  non-extremal rotating BTZ black hole with compact Euclidean time direction at the asymptotic boundary. The value of $h$ restricts the solution to a patch of the black hole spacetime below $r = \sqrt{h}$. The deformed energy for generic values of $h$ is
\begin{align}
	E = {h \over 4 G}\left( 1 - \sqrt{ s(1 - 8G M/h  +   16 G^2 J^2/h^2)}\right) = {r_c^2 \over 4 G}\left( 1 - \sqrt{ s{(r_c^2 - r_+^2)(r_c^2 - r_-^2)\over r_c^4}}\right)
\end{align}
with different values of $s$ depending on $h = r_c^2$. The three possibilities are:
\begin{itemize}
	\item $r_+ < r_c < \infty$: {\bf Euclidean BTZ black hole}\\
	The solution is given by the Euclidean BTZ black hole confined between the tip of the cigar at $r = r_+$ and the finite cutoff surface at $r = r_c = \sqrt{h}$. The deformed energy is 
		\begin{align}
	E = {h \over 4 G}\left( 1 - \sqrt{ 1 - 8G M/h  +   16 G^2 J^2/h^2}\right)
	\end{align}
	\item $r_-<r_c < r_+$: {\bf Intermediate interior of BTZ} \\
	The deformed energy naively complexifies in this range, but the gravity path integral avoids this by assigning a timelike normal to the boundary setting $s = -1$. The deformed energy then reads 
		\begin{align}
	E = {h \over 4 G}\left( 1 - \sqrt{ -(1 - 8G M/h  +   16 G^2 J^2/h^2})\right).
	\end{align}
	  The metric solution with timelike boundary normal is given by 
	  \begin{align}
	  	ds^2 = - |f(r)|dt^2  - {dr^2 \over |f(r)|} + r^2 \left(d\theta-\frac{4iGJ}{r^2}d\tau\right)^2.
	  \end{align}
	   Note that the metric has $(-,-, +)$ signature. The periodicity along the $t$ direction is imposed by  regularity at the inner horizon at $r = r_-$, and will be different than the periodicity in the exterior; the inner horizon and outer horizon have different temperatures. Analytically continuing the timelike direction $t \rightarrow i t$, the metric is just the Lorentzian metric between the inner and outerhorizons of the BTZ black hole. 
       It is interesting to note that the boundary conditions assign $K_{\theta \theta} < 0$ and $\partial_t h = 0$, or equivalently $\partial_\pm \mathrm{Area}(t,r_c)<0$, signifying that surfaces all along the cutoff boundary are trapped surfaces; the boundary conditions impose that we are inside of a black hole.
	\item $0< r_c < r_-$: {\bf Deep interior of BTZ} \\
	Pushing $r_c = \sqrt{h}$ below $r_-$ flips the normal of the boundary to be spacelike again and sets $s = +1$ resulting in a deformed energy identical to the exterior $r_+ <r_c$
	\begin{align}
	E = {h \over 4 G}\left( 1 - \sqrt{ 1 - 8G M/h  +   16 G^2 J^2/h^2}\right).
	\end{align}
The metric is Euclidean and the spacetime is bounded by $r = \sqrt{h}$ and $r = 0$ singularity. Just like the conical defect, the singularity corresponds to an accumulation point $-|J|$. It is interesting that we observe this behavior despite $r = 0$ not actually being singular; $r = 0$ is referred to as a singularity in order to exclude the region $r <0$ that contains closed timelike curves. The accumulation of the flow provides an independent reason to think of $r = 0$ as a singularity.\footnotemark
\end{itemize}
\footnotetext{This form of the deformed energy with $r_c< r_-$ has been previously associated with the region behind the inner horizon in \cite{Guica:2019nzm}.}


\begin{figure}[h!]
   \centering
   \includegraphics[width=0.73\linewidth]{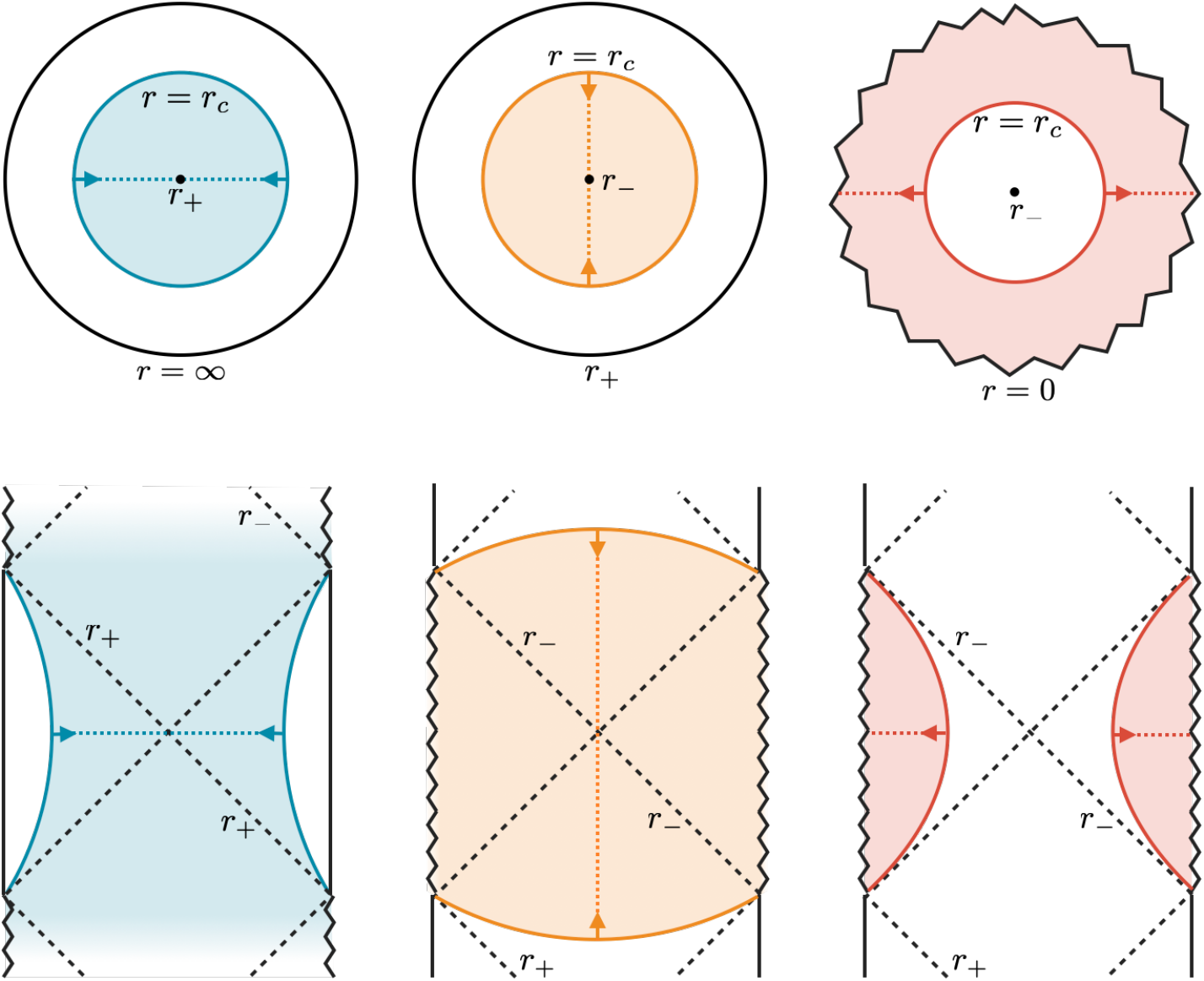} \\
   \vspace{.5cm} \hspace{.25cm}
   \includegraphics[width=0.56\linewidth]{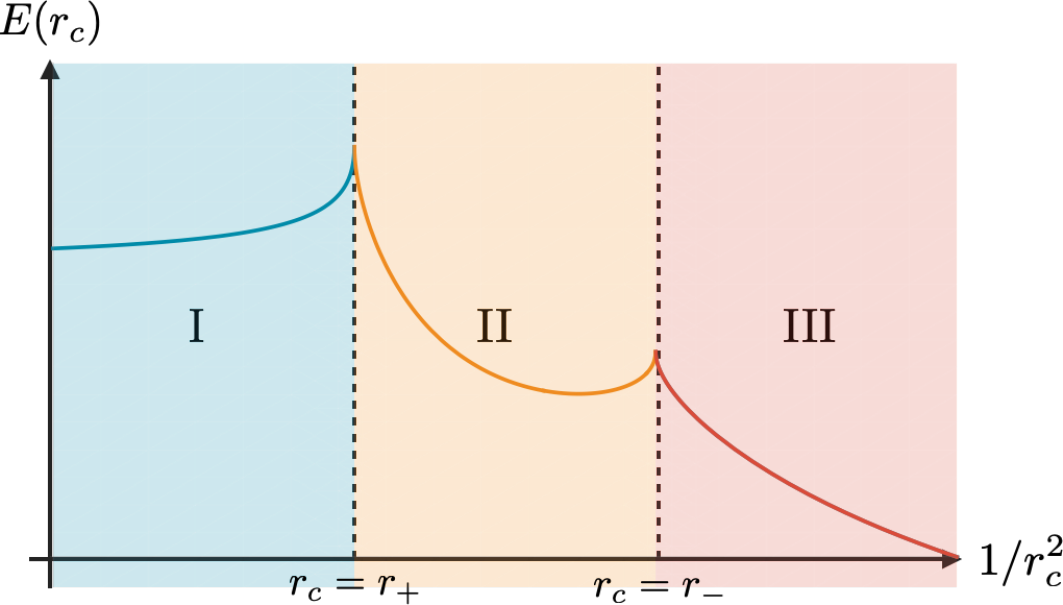}
   \caption{The Euclidean bulk saddles of the gravitation path integral (top row), and their Lorentzian continuation (middle row) defined by fixing the ADM mass \eqref{eq:ADMansatz}, angular momenta \eqref{eq:Jansatz} and the quasi-local energy \eqref{eq:BYansatz} on the cutoff boundary. The arrows indicate the inward pointing normal vector. We list here the resulting bulk solutions that corresponds to fixing the cutoff surfaces at different radius $r_c$. From left to right: (a) exterior ($r_c>r_+$), (b) intermediate interior ($r_-<r_c<r_+$), (c) deep interior ($r_c<r_-$). We also include the flow of deformed energy in each regime in the bottom row.}
   \label{fig:GPI}
\end{figure}
\clearpage

\begin{center}
{ \bf Non-rotating BTZ}
\end{center}
The case of the  non-rotating BTZ black hole is the $J \rightarrow 0$ limit of the previous one. However, since the causal structure of the spacetime is modified at $J = 0$, there will only be two possibilities:
\begin{itemize}
	\item $r_h < r_c < \infty$: {\bf Euclidean $J = 0$ BTZ black hole}\\
	The solution is again the Euclidean BTZ black hole cigar cutoff at $r = r_c = \sqrt{h}$. The deformed energy is 
		\begin{align}
	E = {h \over 4 G}\left( 1 - \sqrt{ 1 - 8G M/h }\right)
	\end{align}
	\item $0<r_c < r_h$: {\bf $J = 0$ BTZ interior} \\
	The flip in the signature of the boundary sets $s = -1$, and the deformed energy becomes 
		\begin{align}
	E = {h \over 4 G}\left( 1 - \sqrt{ -1 + 8G M/h } \right).
	\end{align}
    This has a vanishing accumulation point that occurs during this stage, and coincides with the approach to the spacelike (and genuine) conical singularity at $r = 0$.
\end{itemize}
\begin{center}
{ \bf Extremal BTZ}
\end{center}
The last case we consider is extremal BTZ where $|J| = M$. This also has two possibilities:
\begin{itemize}
	\item $r_h < r_c < \infty$: {\bf Euclidean Extremal BTZ black hole}\\
	The solution is given by the infinite Euclidean BTZ  cigar cutoff at $r = r_c = \sqrt{h}$. Despite being infinitely far away, the horizon can still be reached at finite value for $r_c$. The deformed energy is 
		\begin{align}
	E = {h \over 4 G}\left( 1 - \sqrt{ (1 - 4G M/h)^2 }\right)   = M,
	\end{align}
    where we took the positive branch of the square root for $h > 4 GM$. The deformed energy notabely does not change.
	\item $0<r_c < r_h$: {\bf Extremal BTZ interior} \\
	Beyond the horizon, the boundary actually remains spacetime and $s = +1$ retained. However, since $h < 4 GM$ our branch prescription for the square root gives a non-trivial flow for the energy
	\begin{align}
	E = {h \over 2 G} - M.
	\end{align}
    This accumulates at $-M$ at $r = 0$.
\end{itemize}

\begin{SCfigure}
    \centering
    \includegraphics[width=0.46\linewidth]{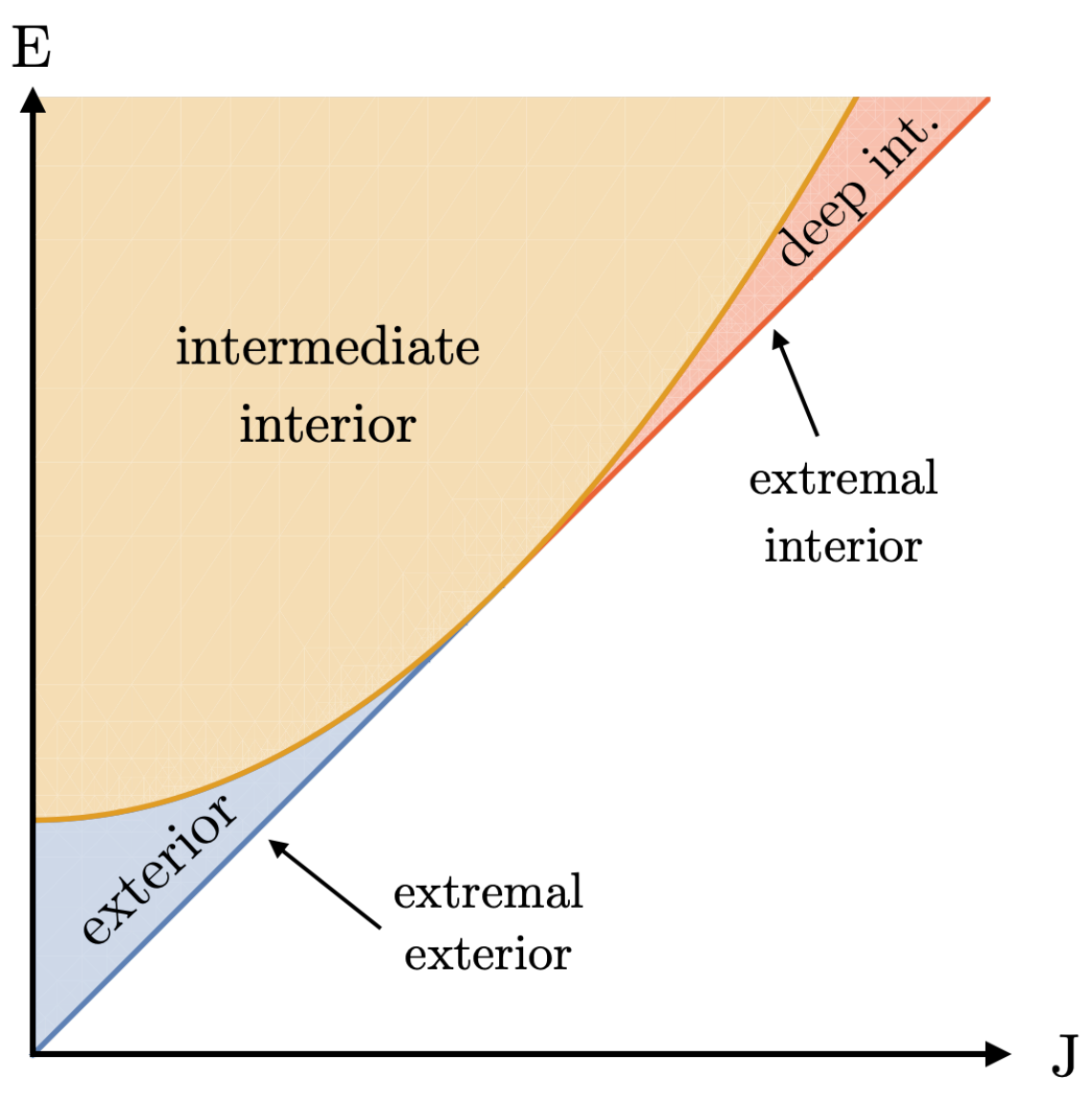}
    \caption{This figure shows the location of the finite cutoff surface as a function of the energy $E$ and angular momentum $J$ for fixed spatial metric component $h$. Unitarity imposes we restrict to $E\ge J$. The interface separating these three regions is the critical energy as a function of $J$ and $h$ defined as the energy where the finite cutoff boundary is at the inner or outer horizon (where the square root in the deformed energy vanishes). This interface intersects the $E = J$ line at exactly one point which moves up/down as $h$ is increased/decreased.}
    \label{fig:phase}
\end{SCfigure}

We remind the reader that all these intricate features are found in the purely boundary analysis of the flow. A summary of the location of the finite cutoff boundary is presented in figure \ref{fig:phase}.


\subsection{Finite cutoff in JT gravity}
\label{sec:GPI-JT}

In this section, we reanalyze the $M>0$ and $J = 0$ case expressed in the variables of JT gravity.This can obtained by dimensional reduction of the transverse spatial coordinate as discussed in \cite{Gross:2019ach}. The metric ansatz is
\begin{align}
	ds^2 = ds^2_{(2d)} + \phi^2 d\theta^2
\end{align}
where $\phi$ is the 2d dilaton and $\phi^2$ plays the role of $h$. The dimensionally reduced theory with fixed BY and asymptotic charges is 
\begin{equation}
S = -\frac{1}{16\pi G}\int_{\mathcal{M}_2}d^2x\sqrt{g^{(2)}}\phi(R^{(2)}+2) + \frac{s}{8\pi G}\int_{\partial \mathcal{M}_2} d\tau\sqrt{\gamma_{\tau\tau}}(\partial_n \phi -\phi K),
\end{equation}
where $\partial_n \equiv \frac{1}{\sqrt{|g(r)|}}\partial_r$ is the outward pointing normalized derivative along the normal (radial) direction.  The 3d ADM mass and the quasi-local energy reduce to
\begin{align}
\label{eq:JT-BC}
M &= \frac{1}{8G} \left( \phi^2 - s(\partial_n\phi)^2\right), \\
E &= \frac{1}{4G}  \left(\phi^2 - \phi\partial_n\phi\right), \label{JTBY}
\end{align}
We see that fixing the tuple $(M,E)$ is equivalent to fixing the dilaton $\phi$ and its derivative, as opposed to the standard Dirichlet boundary condition of fixing $\phi$ and $g_{uu}$.
This particular choice of boundary condition is known to define a microcanonical ensemble in the bulk path integral \cite{Goel:2020yxl}, and here we have shown that it descends from the 3d microcanonical ensemble.

It is worth remarking on an apparent discrepancy between our definition of the Brown-York energy and what appears in the literature. As written in \cite{Maldacena:2016upp} for example, the Brown-York energy is given by
\begin{align}
	\widetilde{E} = {\sqrt{h_{\tau \tau}} \over 4 G}\left(\phi - \partial_n\phi\right),
\end{align}
and where the ADM energy is the asymptotic boundary limit of this expression. The discrepancy between this definition and \eqref{JTBY} suggests a discrepancy in the definition of boundary time. Indeed, implicit in \cite{Maldacena:2016upp} is that the uplifted induced metric on the boundary is  $\gamma_{\tau \tau} d \tau^2 + \phi^2 d\theta^2$ instead  $\phi^2 d\tilde{\tau}^2 + \phi^2 d\theta^2$. Of course, the distinction is moot if  $h_{\tau \tau} = \phi^2$ at the boundary.

Divorcing JT gravity from 3d gravity results in odd behavior of the flowed Brown-York energy. In particular, the form of the deformed energy behind the horizon, where $s = -1$ and $\phi^2 < M$, is
\begin{align}
	E = {\phi^2 \over 4 G} \left( 1 - \sqrt{-1 + 8 G M/\phi^2 }\right).
\end{align}
The issue is that the flowed energies accumulate at $E = 0$ at $\phi = 0$, which doesn't have any significance from the perspective of the JT gravity theory; black hole solutions have no singularity at $\phi = 0$. However, note that $\phi_0$, a constant shift of the dilaton, was not generated by the dimensional reduction. Therefore, in this definition of JT gravity it is the value of $\phi$ that suppresses higher topology contributions, and therefore the accumulation point in $E$ coincides with this region where higher topologies become important.
\subsection{Canonical ensemble partition function}
\label{sec:GPI-can}
We discussed the micro-canonical ensemble gravitational path integral labeled by $(M,J,\Omega)$ and analyzed the saddle points as a function of those parameters. In this section we will define a canonical gravitational path integral and discuss its implication on the boundary canonical partition function. The two ingredients we need to specify are the form of the action including boundary terms as well as the boundary conditions. Our aim is for a definition that makes use all the regions of a black hole.

We start by discussing the boundary conditions. The boundary conditions will be Dirichlet for all the components of the metric, but the question we would like to investigate is: on which surface? One answer is to impose the Dirichlet conditions on the finite cutoff surface. However, this choice would restrict to a single region in the black hole spacetime. This is problematic since the boundary partition function, defined as
\begin{align}
    \sum_{\ \  \   \ \quad E_0 = E_\mathrm{Ground}}^\infty \! \!\!\! \!\!\! \!\!\!  e^{-\beta E_\lambda^{+}(E_0,\lambda)}
\end{align}
where the deformed energy $E^{+}_\lambda(E_0, \lambda) = {1 \over 4 \lambda}\left( 1 - \sqrt{s(1 - 8 \lambda E_0 ) }\right)$, would eventually go complex once the energies complexify. For simplicity, we restricted to the $J = 0$ sector.

To avoid this, we need to allow for the signature of the boundary metric to flip in order to get a contribution from the $s = -1$ solution to the deformed energy. This is like imposing  Dirichlet conditions on the original asymptotic boundary while allowing the signature of the finite cutoff metric to fluctuate. In this case, the partition function takes the form
\begin{align}
	 \sum_{\ \  \   \ \quad E_0 = E_\mathrm{Ground}}^\infty  \! \!\!\! \!\!\! \! \left( e^{-\beta E_\lambda^{+}(E_0,\lambda)}\Theta(E_c - E_0) + e^{+\beta E_\lambda^{-}(E_0,\lambda)}\Theta(E_0 - E_c)  \right).
\end{align}
This allows the inclusion of all energies while maintaining a real partition function. This expression follows from the bulk path integral
\begin{align}
    \int \! {\cal D}g \!\sum_{s = \pm} e^{-S_\mathrm{EH} - {s }   S_\mathrm{BY} },
\end{align}
with Dirichlet boundary conditions on the finite cutoff surface with signatures set by $s$. The story doesn't change by much if we include the sum over $J \neq 0$. Interestingly, for any fixed $J$ the $s = +1$ contribution comes from either the exterior or the deep interior and never both. This can be seen by looking along a vertical slice of the phase diagram Fig.~\ref{fig:phase}.

The negative length of the boundary makes it seem like the partition function will diverge from the lack of suppression of  arbitrarily high energy states. This is not necessarily so. Focusing on the part of the spectrum much larger than $E_c$, we can approximate the partition function as
\begin{align}
    \int dE_0 \rho(E)e^{+ {\beta \over 4 \lambda}\left( 1 - \sqrt{  8 \lambda E_0 -1}\right)} \approx \int dE_0 e^{ \left(2 \pi \sqrt{c \over 3} - {\beta \over \sqrt{2 \lambda}} \right)\sqrt{E_0} },
\end{align}
where the Cardy formula for the density of states was used. We observe that the partition function remains finite for $\beta > 2 \pi \sqrt{2 c \lambda/3}$. For this to happen, it is essential that $E_\lambda$ goes negative for sufficiently large $E_0$.

This partition function diverges for sufficiently large temperature in a way reminiscent of Hagedorn growth in QCD and string theory \cite{Hagedorn:1965st,Atick:1988si,Sundborg:1999ue}. To see this, we change variables from the undeformed to the deformed energy using $|E_\lambda| \approx \sqrt{ E_0 \over  2 \lambda}$ at large $E_0$. The integral then becomes
\begin{align}
    \int d|E_\lambda| e^{ |E_\lambda| \left(2 \pi\sqrt{2c\lambda/3 } - {1 \over T} \right) },
\end{align}
where $e^{|E_\lambda| \sqrt{2 \lambda}}$ plays the role of the stringy density of states with $\sqrt{\alpha'} \sim \sqrt{\lambda}$ and with Hagedorn temperature $T_H =  \sqrt{3/2c \lambda}/2 \pi$.\footnote{See reference \cite{Cooper:2013ffa} for a discussion on \ttbar deformed energies and Hagedorn behavior.}

It is worthwhile to ask if this divergence can be associated to a location in the black hole spacetime. We first note that the high energy behavior of the partition function is universal in any fixed $J$ sector since the deformed energy is universal at large $E_0$ and tends to $ -\sqrt{E_0/2 \lambda}$. If we work in the usual BTZ coordinates, then it is not hard to see that the proper distance from $r = r_c$ to $r_-$ (or to $r=0$ when $J = 0$) goes to zero in the large mass limit. This is surprising; it suggests a resemblance between the singularity and the inner horizon, with possible relevance to the potential instability of inner horizon \cite{Simpson:1973ua,Brady:1995un,Ori:1991zz} and strong cosmic censorship.

\section{Summary + discussion}

\subsection{Summary}

In this paper, we analyzed an extension of the \ttbar flow that includes a deformation dependent cosmological constant that we implemented in tandem with the standard \ttbar flow. We gave a prescription for the multi-step flow to arbitrarily large values of the deformation parameter that avoids complexification of the deformed energies. We pointed out how the metric switches signature between different segments of the flow. See Fig.~\eqref{fig:flow} for a summary diagram.

In the holographic context, we established a relation between the presence or absence of the boundary cosmological constant to the signature of the normal to the holographic boundary being timelike or spacelike, respectively. We showed that the specific sequence of deformations analyzed on the boundary corresponds to flowing the holographic boundary throughout the entire bulk geometry, going past event horizons and reaching arbitrarily close to the black hole singularity.

In the bulk, we defined gravitational path integrals corresponding to deformed canonical and micro-canonical boundary ensembles. For the latter, the gravitational path integral fixes the values of the initial ADM charges corresponding to the dual ``seed'' or undeformed energy and angular momentum, and also fixes the induced spatial metric on the finite cutoff boundary. The renormalization factor between this induced metric and the physical spatial metric of the boundary theory played the role of the deformation parameter controlling the location of the finite cutoff surface in the bulk. Assuming translationally symmetric data on the finite cutoff surface, we defined the ADM charges in terms of local data on this surface that do not depend on its location. This allowed the renormalization factor and the charges to be independently tunable. The renormalization factor parameterizes the boundary theory and the charges specify the ensemble in that theory. We demonstrated that for different ranges of these parameters, the gravity path integral automatically implements the multi-step flow analyzed on the boundary, and defines a path integral anchored to a surface outside or inside black holes.

We used the micro-canonical path integral as a clue for defining the canonical path integral. We showed that the naive definition of simply imposing one set of Dirichlet conditions on the finite cutoff surface for all values of the integration contour leads to a complex boundary partition function. Instead, we sum over all Dirichlet conditions on the finite cutoff boundary that are consistent with the Dirichlet conditions on the asymptotic boundary in the infinite renormalization factor limit (vanishing deformation parameter). This requires the sum over different signatures of the finite cutoff metric depending on the location along the integration contour; it is determined by the specified renormalization factor and the value of ADM charges being integrated over in the path integral. The resulting dual partition function is real and given by the Boltzmann weighted sum over all energies, split up into contributions from different regions of the dual black hole spacetimes. It would be interesting to find the analogous mechanism purely from the boundary perspective. Notably, we found this partition to have Hagedorn growth at temperatures $\sim \sqrt{c \lambda}$ coming from states whose finite cutoff location resides near the inner horizon of a rotating black hole or at the singularity for the non-rotating case.  We conclude with a few remarks and future research directions.

\subsection{Correlation functions of gravitons and matter}

A natural extension of our work is to compute correlation functions anchored to the finite cutoff. Such correlators would provide more fine grained information about the region near the finite cutoff surface, wherever it may be, giving access to parts of the bulk outside of causal contact from the asymptotic boundary.

Stress tensor correlation functions are the most direct type of correlators that can be flowed along with the finite cutoff boundary throughout the bulk. The \ttbar deformation and its extension analyzed in this paper are sufficient to uniquely fix the form of these correlation functions. One has to determine the analogue of the sequence of flow equations but for these correlation functions. The resulting correlators are those of gravitons sourced at the finite cutoff boundary which, in three dimensions, remain localized at the boundary.

For correlation functions of matter operators, more work has to be done. As discussed in \cite{Guica:2019nzm}, a source turned on at the asymptotic boundary does not flow under the standard \ttbar flow. We provide an independent demonstration of this fact in appendix \ref{appendix: matter} where we show that correlators of light matter operators in a \ttbar deformed theory remain anchored to the asymptotic boundary. 
To reproduce finite-cutoff correlators, one also needs to alert the matter sector about the deformation. One way of doing this is by deforming the matter theory in tandem with \ttbar by a matter double trace operator along with the assumption of large-$N$-factorization \cite{Hartman:2018tkw}. This is seen to implement Dirichlet boundary conditions on the finite-cutoff boundary in the bulk. We expect this to continue to work with the addition of the boundary $\Lambda_2$ and pushing the finite cutoff boundary behind the horizon. We leave the analysis of deformed matter and stress tensor correlation functions to future work.
\subsection{A finite interface path integral}

The gravitational path integral we defined treated the finite cutoff surface as a boundary of spacetime. We consider here how to fill in spacetime on both sides of the surface, treating it as an interface rather than a boundary.

Our finite cutoff path integral was represented by the  minus branch ($m = -1$) path integral in \eqref{Zminus}. We noted that  the other branch ($m = +1$) has no saddles without introducing an additional boundary component. Suppose we include it. Then one definition of a finite interface gravitational path integral is simply
\begin{align}
    Z_I = Z_-\times Z_+,
\end{align}
where the $\pm$ refer to the two branches and where there is an additional boundary term implicit in $Z_+$. We assume that the same boundary conditions are fixed on either side of the interface, up to the appropriate sign choices when picking the normal directions. This path integral is naively a product because the interface is codimension 1 and splits the manifold in two. However, the gravitational path integral can spoil this factorization using wormholes.

We can think of this path integral as treating the finite cutoff boundary as an ``observer'' inside the spacetime, although usually additional degrees of freedom are included along the world line (or volume in this case). See for example recent work \cite{Abdalla:2025gzn, Harlow:2025pvj}. There is perhaps more to be said about this connection between \ttbar and observer-centric gravitational path integrals.
\subsection{Entropy and microstates of the singularity}

The micro-canonical gravitational path integral is sensitive to the number of states of the black hole at the chosen asymptotic charges, and should be independent of the location of the finite cutoff surface. In particular, $Z_\mu = \rho(E)$. Indeed, consider the solution for the $J = 0$ case which reads
\begin{align}
    \phi = \sqrt{8GM} \cosh[\sqrt{s}\, r], \quad ds^2  = s dr^2 + \sinh^2[\sqrt{s} \, r] d\theta^2. \label{J0_soln}
\end{align}
Evaluating the on-shell microcanonical particion function reads
\begin{align}
    Z = e^{2 \pi \sqrt{M/2G}}, \label{micro_onshell_partition}
\end{align}
which is consistent with large energy and $J = 0$ density of states for 2 dimensional CFTs with central charge $c = 3/2 G$.

The interesting thing about this result is that it is independent of choice of $s = \pm 1$! This raises a puzzle: in the exterior, we understand that the Bekenstein Hawking entropy, essentially $\ln Z$, is generated in the Gibbons Hawking method due to the capping off at the tip of the cigar
\begin{align}
    \includegraphics[width=0.2\linewidth]{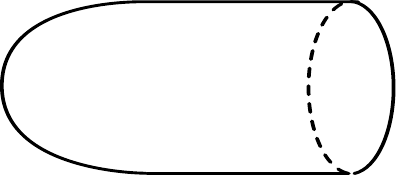}   \nonumber
\end{align}
However, saddle point configuration \eqref{J0_soln} for $s = -1$ is not a cigar, but a cylinder bounded by the finite cutoff surface on one end and the $r = 0$ conical singularity on the other. The manifold is $I \times S^1$, and therefore there is no tip to impose a regularity condition. Therefore a naive application of the Gibbons Hawking method should give an on-shell action linear in size of the $S^1$ resulting in a vanishing entropy ($S = (\beta \partial_\beta - 1) \ln Z$). This same conclusion holds for the $J \neq 0$ case as well.

A way out of this puzzle is to note that regularity at the tip is responsible for producing only the geometric part of the entropy. The Gibbons-Hawking method also generates the matter contribution from the number of matter states running around the $S^1$. There is no matter in this set up, but there is the singularity. We can get the correct result if we suppose that the number of states of the singularity accounts for the Bekenstein-Hawking entropy as seen from the outside
\begin{align}
    \includegraphics[width=0.2\linewidth]{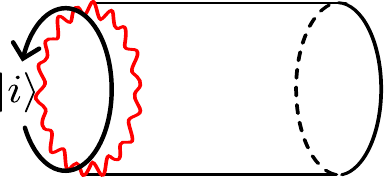} \nonumber.
\end{align}

\subsection{The deformed spectral form factor and interior wormholes}

It would be interesting to find other boundary quantities that are sensitive to the interior. One such quantity is the spectral form factor \cite{Cotler:2016fpe}. Starting with the deformed canonical ensemble partition function (with  $J = 0$ for simplicity)
\begin{align}
    Z_\beta^\lambda = \Tr[e^{-\beta H_\lambda^+} \Theta(E_c - H) + e^{+\beta H_\lambda^-} \Theta(H - E_c)],
\end{align}
it is straightforward to define the deformed spectral factor to be $| Z_{\beta+i T}^\lambda|^2$. This can be organized as
\begin{align}
    |\Tr[e^{-(\beta + i T) H_\lambda^+} \Theta(E_c - H)]|^2 + |\Tr[e^{+(\beta + i T) H_\lambda^-} \Theta(H - E_c)]|^2 + \mathrm{cross \ terms}.
\end{align}
We are interested in nonperturbative gravitational contributions to all these terms. Since the cross terms are the product of terms in different energy ranges, we expect they would vanish under averaging in a small time window or over an ensemble of boundary theories, and therefore do not expect to receive gravitational contributions.\footnote{Although they might have half-wormhole contributions\cite{Saad:2021rcu}.} 

The first and second terms we will think of as exterior and interior contributions to the spectral form factor, respectively. This interpretation follows from the dependence of the location of the finite cutoff boundary on the parameters $(E, \lambda)$. We note that for $J \neq 0$, there will be separate contributions from the three bulk regions divided by the inner and outer horizons.

Analyzing the interior contribution on the bulk and boundary sides will give an additional check on the our proposal of placing the boundary in the interior. On the bulk side, we expect it to receive disconnected and connected contributions of the form
\begin{align}
   {\includegraphics[width=0.8\linewidth]{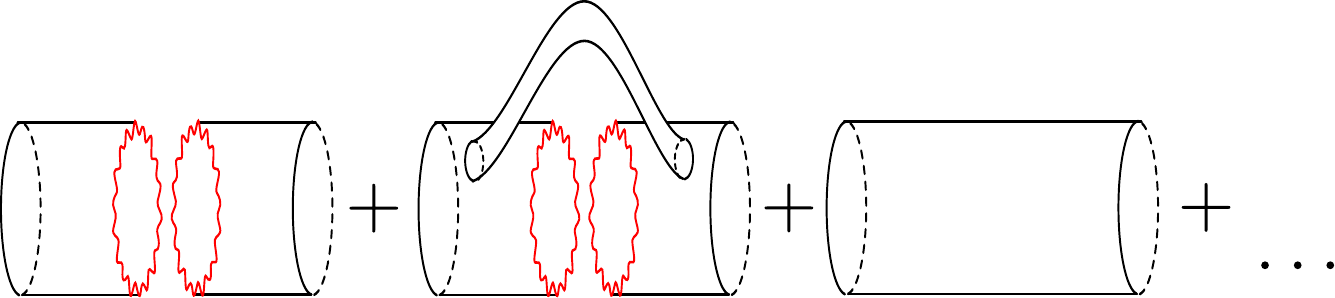}} \nonumber
\end{align}
where all these spacetimes are anchored to the finite cutoff surface in the interior. Note that the spectral statistics in terms of the undeformed energies is unchanged, and we can continue to use those results from, say, JT gravity \cite{Saad:2019lba}. Another line of analysis would be to implement this deformed spectral form factor numerically in SYK. We leave this for future work.

\section*{Acknowledgements}

We thank Shoaib Akhtar, Thomas Faulkner, Marc S.~ Klinger, Adam Levine, Edward Mazenc, Gautam Satishchandran, Ronak Sonni, Eva Silverstein, Ayngaran Thavanesan, Aron Wall, Wayne Weng for insightful discussion. All authors were fasting during the composition of this paper.
\appendix

\section{Insensitivity of probe matter correlation functions to the deformation} \label{appendix: matter}
Before computing the deformed two-point function under standard \ttbar,  we review its computation in the canonical ensemble of the undeformed theory~\cite{Mertens:2022irh}. Using a complete set of energy eigenstates, we have
\begin{equation}
     \braket{O_{1} O_{2}} =  \int dE_{1} dE_{2}\rho_{1}\rho_{2} e^{-\beta E_{2} - \tau (E_{ 2}-E_{1})} |O_{12}|^{2}, 
\end{equation}
where the density of states and matrix elements are
\begin{align}
    \rho_{i} &:= \frac{C}{2\pi^{2}}e^{S_{0}} \sinh 2\pi \sqrt{2 CE_{i}}, \\
    |O_{12}|^{2} &= 2e^{-S_{0}} \frac{\Gamma(\Delta \pm i \sqrt{2 C E_{1}} \pm i \sqrt{2 C E_{2}})}{(2C)^{2\Delta}\Gamma(2\Delta)}.
\end{align}
One can compute this in the semiclassical limit by writing $E_{1} = E_{2} + \omega$, where $E_{1,2}$ are taken to be large (order $C$) and $\omega$ is small (order $1$). This allows us to turn the two-point function into
\begin{equation}
    \braket{O_{1} O_{2}} = \int dE_{2} f(E_{2}) \times \alpha,
\end{equation}
where $f(E_{2}) = e^{2 \pi \sqrt{2E_{2}} - \beta E_{2}}$ and $\alpha$ is the remaining integral over $\omega$ and will be order one. We stress that there is a residual Boltzmann factor $e^{-\tau (E_{2}-E_{1})} = e^{-\tau \omega}$ in $\alpha$. Computing the $E_{2}$ integral via saddle-point gives $\tilde{E}_{2} = \frac{2\pi^{2}}{\beta^{2}}$, and the remaining integral can be recognized as a Fourier transform of the usual conformal two-point function. In other words, the two-point function will be 
\begin{equation}
    \braket{O_{1} O_{2}} \approx_{\rm semiclassical} Z(\beta) \left( \frac{\pi}{\beta \sin \frac{\pi \tau}{\beta}}\right)^{2\Delta}
\end{equation}
We now turn to the deformed two-point function. Note that the matrix elements of matter operators do not get deformed, which means the only difference from the previous computation is that the energies appearing in the Boltzmann factors become the deformed energies
\begin{equation}
     \braket{O_{1} O_{2}}_{\lambda} =  \int dE_{1} dE_{2}\rho_{1}\rho_{2} e^{-\beta E^{\lambda}_{2} - \tau (E^{\lambda}_{ 2}-E^{\lambda}_{1})} |O_{12}|^{2}.
\end{equation}
Repeating the same decomposition and expansion, we find
\begin{equation}
    \braket{O_{1} O_{2}}_{\lambda} = \int dE_{2} e^{2 \pi \sqrt{2E_{2}} - \beta E^{\lambda}_{2}}\alpha_{\lambda}.
\end{equation}
The saddle now occurs at a deformed saddle energy
\begin{equation}
    \tilde{E}^{\lambda}_{2} = \frac{2\pi^{2}}{\beta^{2} + 16 \pi^{2} \lambda},
\end{equation}
and $\alpha_{\lambda}$ is the same as $\alpha$ except for the time-dependent Boltzmann factor we stressed earlier now given by $e^{ \frac{\tau \omega}{\sqrt{1- 8 E_{2}\lambda}}}$.

We can solve for the original $\beta$ above as
\begin{equation}
    \beta = \sqrt{\left(\frac{2\pi^{2}}{\tilde{E}^{\lambda}_{2}}\right)\left(1 - 8\lambda \tilde{E}^{\lambda}_{2}\right)}:= \beta_{\lambda} \sqrt{1 - 8 \tilde{E}_{2}^{\lambda} \lambda}.
\end{equation}
We see that the deformed two-point function will be obtainable from the undeformed one under the identifications
\begin{equation}
    \tau \to \frac{\tau}{\sqrt{1- 8 E_{2} \lambda}}, \quad \beta \to \frac{\beta}{\sqrt{1- 8 E_{2} \lambda}}, \label{fake_rescaling}
\end{equation}
which yields
\begin{equation}
    \braket{O_{1} O_{2}}_{\lambda} \approx_{\rm semiclassical} Z(\beta_{\lambda}) \left( \frac{\pi}{\beta_{\lambda} \sin \frac{\pi \tau}{\beta}}\right)^{2\Delta}.
\end{equation}
Notice that only the normalization flows but not the functional form of the two-point function, in other words, the argument of the sine is the same as in the undeformed theory. The transformation \eqref{fake_rescaling} should be understood as a rescaling to the coordinates of $\lambda = 0$ theory, in which the (normalized) two point function attains the undeformed standard form.\footnote{This is reminiscent of the fake disk discussion in large $p$ SYK analyzed in \cite{Lin:2023trc}, hinting at a possible connection to \ttbar. We leave this to future work.} It is in this sense that the two point function is insensitive to the \ttbar deformation and so the matter theory does not see the finite cut-off surface.

\section{Conservation of bulk charges}
\label{App:conservation}
We demonstrate here that the bulk charges defined in \eqref{eq:ADMansatz}
 and \eqref{eq:Jansatz} is indeed conserved. The idea is to consider the ADM decomposition of $\mathcal{M}$ by constant time surfaces $\Sigma$ \footnote{Note that we choose to foliate the spacetime by constant $\tau$ surfaces here, as opposed to in Sec.~\ref{sec:bulkflow} and Sec.~\ref{sec:GPI} where $\Sigma$ is taken to be a constant radial slice.}. The conservation of charges then follows from the bulk Hamiltonian and momentum constraints
\begin{align}
    \mathcal{H} &= \frac{1}{2}\left( R_\Sigma + K^2 - K_{ab}K^{ab} + 2s \right) = 0, \\
    \mathcal{H}_b &= 2D^a(K_{ab}-h_{ab}K)=0,
\end{align}
where $K_{ab}$ and $R_\Sigma$ and $D^a$ are the extrinsic curvature, scalar curvature, and covariant derivative on $\Sigma$ and $s=n^an_a=\pm 1$ is the signature of $\Sigma$. 
In terms of our parametrization of the metric,
\begin{equation}
    ds^2 = f(r)d\tau^2 + g(r)dr^2 + h(r) (d\theta + N_\theta(r) d\tau)^2,
\end{equation}
these constraints read
\begin{align}
    \mathcal{H} &= \frac{s}{2}\left( -\frac{2h(r) N'_\theta{}^2(r)}{f(r)g(r)} - \frac{h''(r)}{g(r)h(r)}+\frac{h(r)'(g(r)h(r))'}{(g(r)h(r))^2}+4\right) = 0, \\
    \mathcal{H}_\theta &= -\sqrt{s g(r) h(r)}\left(\frac{h^{3/2}(r)N'_\theta(r)}{\sqrt{f(r)g(r)}}\right)'=i\sqrt{sg(r)h(r)}J'(r) = 0, \\
    \mathcal{H}_r &= 0.
\end{align}
We immediately see that one of the momentum constraint implies that $\partial_r J(r)=0$.
For the mass, we can identify that
\begin{align}
    \partial_r M(r) &= \frac{1}{2}\partial_r \left( -\frac{h^2(r)N'_\theta{}^2(r)}{4f(r)g(r)}-\frac{h'{}^2(r)}{4g(r)h(r)}+h(r) \right) \\
    &=\frac{1}{2}\partial_r \left( \frac{J^2(r)}{4h(r)}-\frac{h'{}^2(r)}{4g(r)h(r)}+h(r) \right) =\frac{sh'(r) \mathcal{H}}{4} = 0,
\end{align}
thus confirming the functional $M$ is indeed conserved.
 
\section{Variation of bulk microcanonical action}
\label{App:variation}
We consider the variation of the following Euclidean action:
\begin{equation}
S = \underbrace{-\frac{1}{16\pi G}\int_\mathcal{M}d^3x\sqrt{g}(R+2)}_{S_{\rm EH}} -
\underbrace{\frac{s}{8\pi G}\int_{\partial\mathcal{M}}d^2x \sqrt{|\gamma|}(K-1)}_{S_{\rm GHY}} + \,S_{\rm micro}
\end{equation}
under the metric ansatz
\begin{equation}
    ds^2 = f(r)d\tau^2+g(r)dr^2+h(r)(d\theta+N_\theta(r) d\tau)^2,
\end{equation}
where
\begin{equation}
    S_{\rm  micro} = \frac{1}{2\pi}\int_{\partial \mathcal{M}}d^2x \sqrt{|\gamma|}\left(\frac{E}{h(r)}+\frac{iN_\theta(r) J}{\sqrt{|f(r)h(r)|}}\right)
\end{equation}
and $s=\pm 1$ depending on the signature of the induced metric $\gamma$. We assume that $h(r)>0$ while $f(r)$ and $g(r)$ can take either positive or negative values.
We fix the triple $(E,J,h)$ on $\partial \mathcal{M}$ as our boundary condition. This is equivalent to fixing $(M,J,E)$ on $\partial \mathcal{M}$ as discussed around Eq.~\eqref{eq:MJEh} and corresponds to a defining a microcanonical ensemble.

The variation of $S_{\rm EH}+S_{\rm GHY}$ gives \cite{Brown:1992br}
\begin{align}
    \delta (S_{\rm EH}+S_{\rm GHY}) = {\rm EOM} - \frac{s}{16\pi G}\int_{\partial \mathcal{M}}d^2x \, \sqrt{|\gamma|} (\gamma^{ab}(K-1) - K^{ab})\delta g_{ab}.
\end{align}
For our metric ansatz, the boundary term evaluates to
\begin{align}
\begin{split}
    &\quad \frac{-s}{16\pi G}\int_{\partial \mathcal{M}}d^2x \, \sqrt{|\gamma|} (\gamma^{ab}(K-1) - K^{ab})\delta g_{ab}\\
    &= \frac{1}{16\pi G}\int_{\partial \mathcal{M}}d^2x \frac{(h'(r)-2h(r)\sqrt{|g(r)|})\delta f+(f'(r)-2f(r)\sqrt{|g(r)|})\delta h - 2h^2(r)N'_\theta(r)\delta N_\theta}{2\sqrt{f(r)g(r)h(r)}} \\
    &=-\frac{1}{4\pi}\int_{\partial \mathcal{M}}d^2x \left( \frac{E}{\sqrt{|f(r)h(r)|}}\delta f+2iJ\delta N_\theta \right),
\end{split}
\end{align}
where we have used the expressions in \eqref{eq:Jansatz} and \eqref{eq:Eansatz}. Recall that we have $\delta h=0$ as one of our boundary conditions.
The variation of $S_{\rm micro}$ reads
\begin{align}
\begin{split}
    \delta S_{\rm micro} &= \frac{1}{2\pi}\delta \int_{\partial \mathcal{M}} d^2x \left(\frac{\sqrt{|f(r)|}}{\sqrt{h(r)}}E + i J N_\theta\right) \\
    &= \frac{1}{4\pi}\int_{\partial \mathcal{M}} d^2x \left(\frac{E}{\sqrt{|f(r)h(r)|}}\delta f + 2iJ\delta N_\theta \right)
\end{split}
\end{align}
since $\delta E = \delta J=0$. Hence we verify that the boundary terms in $\delta S_{\rm EH}+\delta S_{\rm BHY}$ and $\delta S_{\rm micro}$ cancels and we indeed have a well-defined variational problem.

\small
\bibliographystyle{ourbst}
\providecommand{\href}[2]{#2}\begingroup\raggedright\endgroup

\end{document}